\documentclass[a4paper,floatfix,superscriptaddress,aps,prl,twocolumn,showpacs,notitlepage,nofootinbib, nobibnotes, longbibliography]{revtex4-2}
\usepackage[utf8]{inputenc}
\usepackage{bbold}
\usepackage{graphicx}

\graphicspath{ {images/} }
\usepackage[inkscapepath=images,inkscape=off]{svg}
\usepackage{braket}
\usepackage[normalem]{ulem}
\usepackage[T1]{fontenc}
\usepackage{lmodern}
\usepackage{textcomp}
\usepackage{gensymb}
\usepackage{amsfonts}
\usepackage{amssymb}
\usepackage{amsmath}
\usepackage{amsthm}
\usepackage{nicefrac}

\newcommand{\figref}[1]{\figurename~\ref{#1}}

\usepackage{mathtools}
\usepackage{microtype} 
\usepackage{dcolumn}
\usepackage{enumitem,color}
\usepackage{dsfont}
\usepackage{siunitx}
\let\svqty\qty
\usepackage{physics}
\let\qty\svqty

\usepackage[caption=false]{subfig}
\usepackage{ragged2e}
\DeclareCaptionJustification{justified}{\justifying}
\usepackage{bbm}
\usepackage{bm}
\usepackage{synttree}
\usepackage{float}
\usepackage{xcolor}
\usepackage{tikz}
\usepackage{tikzsymbols}

\usepackage{booktabs}
\usepackage{upgreek, eurosym}
\usepackage[colorlinks=true]{hyperref}

\graphicspath{{images/}}

\newcommand{\mac}{M\!AC}

\usepackage[nolist]{acronym}

\begin{document}
\usetikzlibrary{arrows.meta}
\tikzsymbolsset{after-symbol={}}
\tikzsymbolsdefinesymbol {hvbas}{ S } {%
    \begin{tikzpicture}[/tikzsymbolsstyle, x=1.1ex,y=1.1ex, line width=0.07ex*\tikzsymbolsscaleabs]
    \draw[{Latex[scale=0.5]}-{Latex[scale=0.5]}] (0,-1) -- (0,1);
    \draw[{Latex[scale=0.5]}-{Latex[scale=0.5]}] (-1,0) -- (1,0);
    \end{tikzpicture}
}
\tikzsymbolsdefinesymbol {adbas}{ S }{%
    \begin{tikzpicture}[/tikzsymbolsstyle, x=1.1ex,y=1.1ex, scale=#1, line width=0.07ex*\tikzsymbolsscaleabs]
    \draw[{Latex[scale=0.5]}-{Latex[scale=0.5]}] (-1*0.707,-1*0.707) -- (1*0.707,1*0.707);
    \draw[{Latex[scale=0.5]}-{Latex[scale=0.5]}] (-1*0.707,1*0.707) -- (1*0.707,-1*0.707);
    \end{tikzpicture}
}

\newacro{gps}[GPS]{Global Positioning System}
\newacro{hmac}[MAC]{Message Authentication Code}
\newacro{itsec}[i.t.-security]{information-theoretic security}
\newacroplural{itsec}[i.t.-secure]{information-theoretic secure}
\newacro{qkd}[QKD]{Quantum Key Distribution}
\newacro{spdc}[SPDC]{Spontaneous Parametric Down Conversion}
\newacro{ttm}[TTM]{Time-Tagging Modules}
\newacro{ttp}[TTP]{Trusted Token Provider}

\title{Demonstration of quantum-digital payments}

\author{Peter Schiansky}
\thanks{These two authors contributed equally.}
\affiliation{University of Vienna, Faculty of Physics, Vienna Center for Quantum Science and Technology (VCQ), 1090 Vienna, Austria}

\author{Julia Kalb}
\thanks{These two authors contributed equally.}
\affiliation{University of Vienna, Faculty of Physics, Vienna Center for Quantum Science and Technology (VCQ), 1090 Vienna, Austria}

\author{Esther Sztatecsny}
\affiliation{University of Vienna, Faculty of Physics, Vienna Center for Quantum Science and Technology (VCQ), 1090 Vienna, Austria}

\author{Marie-Christine Roehsner}
\thanks{Current address: QuTech \& Kavli Institute of Nanoscience, Delft University of Technology, 2628 CJ Delft, The Netherlands.}
\affiliation{University of Vienna, Faculty of Physics, Vienna Center for Quantum Science and Technology (VCQ), 1090 Vienna, Austria}

\author{Tobias Guggemos}
\affiliation{University of Vienna, Faculty of Physics, Vienna Center for Quantum Science and Technology (VCQ), 1090 Vienna, Austria}

\author{Alessandro Trenti}
\thanks{Current address: Security and Communication Technologies, Center for Digital Safety and Security, AIT Austrian Institute of Technology GmbH, Giefinggasse 4, 1210 Vienna, Austria.}
\affiliation{University of Vienna, Faculty of Physics, Vienna Center for Quantum Science and Technology (VCQ), 1090 Vienna, Austria}

\author{Mathieu Bozzio}\email[Corresponding author: ]{mathieu.bozzio@univie.ac.at}
\affiliation{University of Vienna, Faculty of Physics, Vienna Center for Quantum Science and Technology (VCQ), 1090 Vienna, Austria}

\author{Philip Walther}\email[Corresponding author: ]{philip.walther@univie.ac.at}
\affiliation{University of Vienna, Faculty of Physics, Vienna Center for Quantum Science and Technology (VCQ), 1090 Vienna, Austria}
\affiliation{Christian Doppler Laboratory for Photonic Quantum Computer, Faculty of Physics, University of Vienna, 1090 Vienna, Austria}

%


\begin{abstract}

 Digital payments have replaced physical banknotes in many aspects of our daily lives. Similarly to banknotes, they should be easy to use, unique, tamper-resistant and untraceable, but additionally withstand digital attackers and data breaches. Current technology substitutes customers' sensitive data by randomized tokens, and secures the payment's uniqueness with a cryptographic function, called a cryptogram. However, computationally powerful attacks violate the security of these functions. Quantum technology comes with the potential to protect even against infinite computational power. Here, we show how quantum light can secure daily digital payments by generating inherently unforgeable quantum cryptograms. We implement the scheme over an urban optical fiber link, and show its robustness to noise and loss-dependent attacks. Unlike previously proposed protocols, our solution does not depend on long-term quantum storage or trusted agents and authenticated channels. It is practical with near-term technology and may herald an era of quantum-enabled security.

\end{abstract}

\maketitle

\noindent The development of quantum algorithms compromising modern cryptography has triggered a global research for stronger security levels \cite{GS:PRL21,MLL:NatPhot12,S:SIAM97}:
the security of current cryptographic schemes relies on computationally hard mathematical problems (known as \textit{computational security}), which should be replaced by quantum-resistant schemes.
While research and standardization for such quantum-resistant solutions is blossoming, some of them have already been broken by computational attacks~\cite{attack-rainbow,attack-sidh,attack-sphincs}.

Quantum-mechanical laws, on the other hand, can provide security against adversaries with unlimited computational power for some tasks \cite{BB84,Pan:RevMod20}. 
This type of security, known as \textit{\ac{itsec}}, is one of the motivations towards a quantum internet~\cite{WEH:Sci18}. 
So far, \ac{qkd} is the most mature and widely implemented quantum technology: it allows two mutually trusted parties to communicate securely over a public channel. \ac{qkd} can already establish \acp{itsec} connections over $500$~km of optical fiber \cite{WYH:NatPhot22,BBR:PRL18} and $1,000$~km of free space using satellites~\cite{Pan:Nature20,BAL:npjqi17}.

In the modern era of digital payments ranging from contactless purchases to online banking, a plethora of new security threats arise. One significant threat occurs when customers interact with untrusted merchants, who may not have sufficient means to protect against external hackers, or may be malicious themselves~\cite{PCIstandard}. 
In that case, a binding commitment between the customer, the merchant and the bank or payment-network is required to guarantee the validity of a transaction. Such a bond usually comes in the form of a cryptogram \cite{corella2014interpreting,emvstandard}, which is the output of a hash function that guarantees the one-time nature of each purchase. Since not all parties involved are trusted, \ac{qkd} is not suitable to provide i.t. security here, and other quantum solutions need to be established. Device-independent versions of QKD \cite{ZLR:Nature22,NDN:Nature22,LZZ:PRL22}, which do not assume trusted quantum sources or detectors, are also inadequate, since the final classical output (i.e. the cryptogram) is handled by the untrusted parties themselves.

Motivated by the no-cloning property of quantum mechanics, previous works have investigated the potentials and drawbacks of using quantum light in the prevention of banknote counterfeiting \cite{Wie:acm83,AC:stoc12,BC+:npjQI17} and double-spending with tokens or credit cards \cite{PY+:pnas12,BOV:npj18,GAA:pra18,BDG:PRA19,HS:NJP20}. Introducing this fundamentally new type of money to everyday scenarios is, however, technologically challenging: quantum states must be stored over days or months to ensure flexible spending.
This is far beyond state-of-the-art quantum storage times, which range from a few microseconds to a few minutes \cite{MMZ:natcomms21,VHC:NC18,HE+:jmo16}. Recently, an interesting alternative was proposed, replacing quantum storage by a network of trusted agents and authenticated channels, positioned at precise spatial locations with respect to the spending points \cite{Kent:PRA20,Kent:npj22}. From a practical standpoint, this approach presents new drawbacks, as customers and online shoppers do not have the means to securely set up complex trust networks for everyday transactions. Furthermore, accurately monitoring the spatial and temporal coordinates of verifiers requires a trusted \ac{gps}, which opens the door to undesired spoofing-type attacks~\cite{spoof11}.

\begin{figure}[t]
    \begin{center}
        \includesvg[width=0.8\linewidth, inkscapelatex=false]{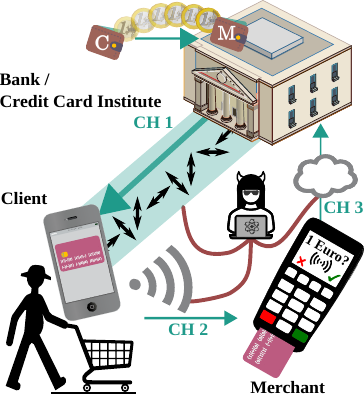}
        \caption{\textbf{Simplified representation of quantum-digital payments.} As in classical payments, we consider three parties: a Client, a Merchant and a Bank/Creditcard institute. In contrast to \cite{Kent:npj22}, we do not assume any quantum or classical communication channel to be trusted (i.e. CH\,1, CH\,2 and CH\,3 are insecure), except an initial prior step between the Bank and Client for an account creation. All parties involved apart from the Bank can also act maliciously. During a payment, the Bank sends a set of quantum states to the Client's device (e.g. phone, computer, etc.), who measures them and transforms them into a quantum-secured payment token -- \textit{cryptogram} -- which we display here as a one-time credit card. The Client uses this classical token for paying at the Merchant, who then contacts the Bank for payment verification. If the payment is accepted, the bank transfers the money from the Client's account to the Merchant's.}
        \label{fig:principles}
    \end{center}
\end{figure}

In this work, we show how quantum light can provide practical security advantages over classical methods in everyday digital payments. As shown in \figref{fig:principles}, we generate and verify i.t.-secure quantum cryptograms, in such a way that the unforgeability and user privacy properties from previous experimental works holds \cite{Kent:npj22}, but all intermediate channels, networks and parties are untrusted, thus significantly loosening the security assumptions. 
Only one authenticated communication (between the client and their payment provider) has to take place at an arbitrary prior point in time. The concealment of the customers' sensitive information is guaranteed by an \acp{itsec} function, and the commitment to the purchase is guaranteed by the laws of quantum mechanics. Additionally, no cross-communication is required to validate the transaction in the case of multiple verifier branches. Our implementation is performed over a $641$m urban fiber link, and can withstand the full spectrum of noise and loss-dependent attacks, including those exploiting reporting strategies~\cite{BCD:PRX21}.

\bigskip

\begin{figure}
    \begin{center}
        \includesvg[width=.5\textwidth]{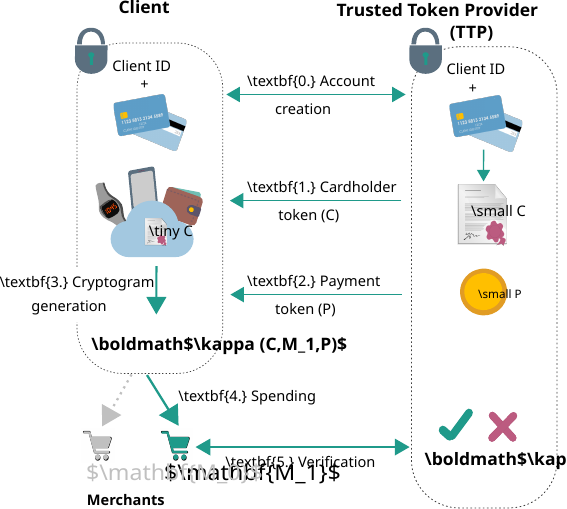}
        \caption{\textbf{Classical digital payments.} \textit{Step $0$}: The Client sets up an account at the \acf{ttp}, 
        providing their secret ID and sensitive credit card information through an authenticated and encrypted channel. \textit{Step $1$}: The Client authenticates with the \acs{ttp}, and requests a cardholder token $C$, which the \acs{ttp} sends through a secure channel. \textit{Step $2$}: The \acs{ttp} randomly generates a one-time token $P$ and sends it to the Client through a secure channel. \textit{Step $3$}: The Client's device uses the stored secret token $C$, the public merchant ID $M_i$, and the payment token $P$ to compute a cryptogram $\kappa\left(C,M_i,P\right)$. \textit{Step $4$}: The Client spends the cryptogram at the chosen Merchant. \textit{Step $5$}: The Merchant verifies the cryptogram with the \acs{ttp}, and accepts or rejects the transaction.}
        \label{fig:classical}
    \end{center}
\end{figure}

\noindent\textbf{Digital payments.} 
We first describe the main security concepts of today's online and contactless purchases~\cite{corella2014interpreting,emvstandard} (actual implementation may vary). 
Following \figref{fig:classical}, each Client initially sets up an account with a \ac{ttp} via a secure communication channel. The \ac{ttp} is usually the Client's bank, credit card provider, or a trusted external company. Through this initial step, the Client is assigned a unique identification token $C$, which is stored securely on both the Client's and \ac{ttp}'s devices. The Client's stored data can be e.g. an electronic wallet or a virtual credit card stored on a smartphone, watch, etc. 

When the Client wishes to purchase goods or services from a given Merchant $M_i$, it has to be ensured that malicious parties, including untrusted Merchants, cannot spend in the Client's name at another place or time. 
That is why the Client receives a one-time payment token $P$ from the Merchant or \ac{ttp}, which is used to compute a \textit{cryptogram}, an output of a function of their secret token $C$, the Merchant's public ID $M_i$, and the one-time payment token $P$. 
We note here that the Merchant ID $M_i$ must be valid and honest (e.g. provided by a Public Key Infrastructure or a securely pre-shared locally stored database).
This cryptogram, which we call $\kappa\left(C,M_i,P\right)$, is communicated to the Merchant, who then sends it to the \ac{ttp} for verification. The \ac{ttp} can verify the signature and uniqueness of the cryptogram, since they have knowledge of all three inputs $C$, $M_i$ and $P$.

In real-world applications, the cryptogram is the output of a cryptographic hash- or encryption function~\cite{Danushka.2017,emvstandard} that is computationally secure.
However, this would allow a malicious party with sufficient computational power to run through all input combinations of $C$, $P$ and $M_i$ until they recover the one combination that matches the original cryptogram. In that case, the Client's ID and payment data are completely compromised. 

\medskip
\noindent\textbf{Quantum advantage.} Considering these attacks only, previous quantum digital signature schemes can provide \ac{itsec} \cite{AWK:PRA16,YYC:NSR22}. However, they typically require \ac{qkd} channels and classical authentication between all three parties.

In this work, we propose a quantum solution that requires only one \ac{qkd} for the initial step between Client and TTP (Step $1$ in \autoref{fig:classical}).  
It is similar to classical digital payments, but replaces the one-time payment token $P$ by a sequence $\ket{P}$ of quantum states.
That is to say, $\kappa\left(C,M_i,P\right)$ becomes $\kappa\left(C,M_i,\ket{P}\right)$ and steps 2-5 from \figref{fig:classical} are modified as follows:

\begin{enumerate}[start=2, left=0pt]
    \item The \ac{ttp} generates a random bitstring $b$ and a random conjugate basis-string $\mathcal{B}$ of length $\lambda$. 
    Each bit $b_j$ is encoded onto a quantum state prepared in $\mathcal{B}_j$, where $j \in \{1;...;\lambda\}$.
    This constitutes the classical description $(b,\mathcal{B})$ of the quantum token $\ket{P}$, which the \ac{ttp} stores under the Client's ID $C_{\textnormal{ID}}$ (e.g. let $\lambda=4$ with the basis $\mathcal{B}_j \in \{+\!/\!-; 0/1$\} such that \mbox{$(b,\mathcal{B})$ = ``0101 0011''} would result in $\ket{P}$ = ``$\ket{+} \ket{-} \ket{0} \ket{1}$'').
    The length $\lambda$ depends on the tolerated success probability of an attack and the number of available merchants. 
    \item Upon receiving $\ket{P}$, the Client chooses the Merchant $M_i$ out of a database that was securely pre-shared with the TTP.
    Next, they calculate \mbox{$m_i = \mac(C, M_i)$}, which is the output tag of an \acp{itsec} \ac{hmac}~\cite{gilbert1974codes, Fak.1979,rosenbaum1993lower, Wegman.1981,Ghosh.2021} that takes the secret token $C$ and the chosen Merchant's public ID $M_i$ as input (see Methods). 
      The Client interprets $m_i$ as a basis-string and privately measures the whole sequence $\ket{P}$ according to $m_i$.
      The resulting string of measurement outcomes $\kappa_i \overset{m_i}{\longleftarrow}\ket{P}$ constitutes the cryptogram.
    \item The Client sends $\kappa_i$ along with their ID $C_{\textnormal{ID}}$ to the Merchant, who forwards this together with its $M_i$ as $\{\kappa_i,M_i, C_{\textnormal{ID}}\}$ to the \ac{ttp} for verification.
    \item To authorize the purchase, the \ac{ttp} looks up $C$ and $(b,\mathcal{B})$, and calculates \mbox{$m_i = \mac(C, M_i)$} for the Client's ID  .
        The \ac{ttp} accepts the transaction if and only if $(\kappa_i)_j = b_j$ when \mbox{$(m_i)_j = \mathcal{B}_j$}.
        The \ac{ttp} rejects otherwise.
        
\end{enumerate}

\noindent The protocol's security depends on the upper bound of the success probability to produce two valid, distinct cryptograms $\kappa_i$ and $\kappa_j$ for two distinct Merchants $M_i$ and $M_j$; we call this $p_d$ (c.f. following two sections).
Another possible attack is to forge an output tag, such that $\acs{hmac}(C, M_i)=\acs{hmac}(C, M_j) \Leftrightarrow m_i=m_j \Leftrightarrow \kappa_i=\kappa_j$; we call the respective probability $p_t$.

In an i.t.-secure MAC,
$p_t = 1/\abs{m} = \abs{M}/\abs{C} = 1/\sqrt{\abs{C}}$, where $\abs{m}$, $\abs{M}$ and $\abs{C}$ refer to the cardinality of the MAC, the Merchant ID and the Client's secret token respectively. Here we assume that $\abs{m} = \abs{M} = \sqrt{|C|}$. Since $p_d$ and $p_t$ should be of the same order of magnitude we choose $p_d \approx p_t = 1/\sqrt{\abs{C}}$. This will yield the number $N$ of quantum states necessary to verify one bit of the cryptogram. As the bit length of any MAC is defined as $\log_2 (\abs{m})$, the entire length of the quantum token will be given by $\lambda = N \cdot \log_2 (\abs{m}) = N \cdot \log_2 (\sqrt{\abs{C}})$.
Any additional parameter that should be committed to during the transaction (e.g., payment amount) can be added as an input to the MAC function.

Just like \ac{qkd} provides \ac{itsec} for key exchanges such as Diffie–Hellman~\cite{Diffie}, our scheme provides \ac{itsec} for the one-time property of cryptograms: while the concealment of $C$ is guaranteed by the \acp{itsec} \ac{hmac}, the commitment to $M_i$ is ensured by the irreversible nature of quantum measurements (see Methods). Notably, our quantum commitment is not limited by the impossibility theorem of quantum bit commitment \cite{LC:PRL97,M:PRL97}, in which one of the two parties can delay their quantum measurements in time. This is because in our protocol one of the interacting parties is assumed to be honest (the \ac{ttp}). 

We note that our implementation contrasts with those of \ac{qkd} schemes in two ways. First, the choice of measurement basis is deterministic as opposed to random. This effectively commits the purchase to a given Client token and Merchant. Second, the measurement bases are never publicly revealed, which has the interesting benefit of hiding the Merchant that was chosen by the Client until verification is required~\cite{Kent:npj22}. 

\medskip

\noindent\textbf{Loss-dependent security.} 
    Although the security of commitment is guaranteed by the laws of quantum mechanics in theory, certain considerations have to be taken into account in a practical setting: 
    
    Due to imperfections of real devices (inaccurate state preparation, lossy quantum channels, non-unit detection efficiency), some quantum states will divert from their ideal classical descriptions, or get lost along the way. 
    In fact, some bits in step 5 will be unequal, although measured in the same basis  (i.e.\, $(\kappa_i)_j \ne b_j$ when \mbox{$(m_i)_j = \mathcal{B}_j$}) and the protocol would abort even though it was followed honestly.
    This is why we have to allow for errors and losses during the verification procedure. 
    In turn, a malicious party can exploit this new allowance to circumvent the commitment or double-spend the cryptogram. 
    
    As an example, assume that the \ac{ttp} tolerates as many as $50\%$ losses. 
    A malicious Client could then measure half of the quantum token $\ket{P}$ in the basis for $M_0$ and the other half in the basis for $M_1$, effectively creating two successfully committed tokens. 
    While double-spending is certainly possible with a loss-rate as high as $50\%$, we use semidefinite programming to identify combinations of error- and loss-rates for which an attack can still be detected.
    Intuitively, the derivation involves searching for the cheating strategy that minimizes the malicious party's introduction of excess errors and losses in the protocol (see Methods).
    We note that, to the best of our knowledge, such powerful loss-dependent attacks were not considered in previous quantum token implementations~\cite{Kent:npj22,BOV:npj18}. 
    
    \begin{figure*}[t]
        \centering
              \includesvg[width=.75\linewidth,inkscapelatex=false,inkscapearea=page]{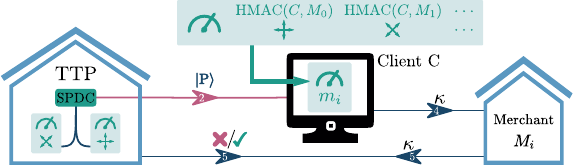}
              \includesvg[width=0.9\linewidth]{images/link-new_v02}
        \caption{\textbf{Experimental quantum-digital payments.} \textbf{a)} The \acf{ttp} creates entangled photon pairs using a \acf{spdc} source. One photon's polarization is randomly measured by the \ac{ttp} in either the linear or diagonal basis, creating the classical description $(b,\mathcal{B})$, which remotely prepares the quantum token $\ket{P}$ on the second photon. The latter is sent to the Client through a $641$m long optical fiber link, who measures its polarization in a basis $m_i = \mac(C, M_i)$ specified by a \acf{hmac} of the Merchant's ID $M_i$ and the Client's private token $C$, and thereby obtains the \textit{cryptogram} that is $\kappa_i \overset{m_i}{\longleftarrow} \ket{P}$. Classical communication between the \ac{ttp}, Client and Merchant is used to verify the compatibility of  $\kappa$, $M_i$ and $C$ with  $(b,\mathcal{B})$. Red (blue) lines indicate quantum (classical) channels. The arrow numbering indicates the steps from \figref{fig:classical}.
		\textbf{b)} Satellite image of the two buildings housing the \ac{ttp}, Client and Merchant. A \SI{641}{\meter} optical fiber link connects the parties.}
		\label{fig:setup}
    \end{figure*}

\medskip
\noindent\textbf{Experimental demonstration.} We implement our quantum-digital payment scheme over the deployed optical fiber link depicted in \figref{fig:setup}. The \ac{ttp} employs a spontaneous parametric down conversion (\ac{spdc}) source to create a pair of polarization-entangled photons in the state \mbox{$\ket{\Psi^-} = \left( \ket{HV} - \ket{VH}\right)/\sqrt{2}$}.
The \ac{ttp} keeps one of these photons and employs a $50/50$ beamsplitter to probabilistically direct it to one of two polarization projection stages, measuring its polarization in either the linear H/V $(\hvbas)$ or diagonal D/A $(\adbas)$ basis. This creates the random classical description $(b,\mathcal{B})$ and remotely imprints the payment token $\ket{P}$ onto the second photon. 
    \par 
    The payment token is sent to the Client, located in another building, through a $641$m optical fiber link. Using a half-wave plate, the Client commits to exactly one Merchant from the set $\{M_0, M_1\}$ by measuring either in the H/V basis for $m_0 = \mac\left(C,M_0\right)$ or in the D/A basis for $m_1 = \mac\left(C,M_1\right)$. 
    In this way, the Client retrieves the classical cryptogram $\kappa\left(C,M_i,\ket{P}\right)$, and forwards it to the Merchant, who is, for convenience, located in the same laboratory. Note that in the case of more than two merchants, the token is split into several sub-tokens that are each measured either in H/V or D/A. We discuss how to adapt the token length in the following section. 
    \par
    At any later time, the Merchant transmits the \textit{cryptogram} received by the Client back to the \ac{ttp}, using a classical channel that links the two buildings. The \ac{ttp} finally checks the compatibility of $(b,\mathcal{B})$ with $M_i$, $C$ and $\kappa_i$, and accepts or rejects the requested transaction. 
    
\noindent\textbf{Results.} We repeatably execute the experiment for both commitments in H/V and D/A. The average measured error rate is $1.45\:\pm 0.01\%$ for H/V and $3.28\:\pm 0.01\%$ for D/A. The overall losses, combining the deployed fiber link and the Client's setup (including detection efficiency), are estimated at $22.40\:\pm 1.50\%$, while the multiphoton emission probability, measured through a correlation measurement, 
is$6.76\:\pm 0.12\%$.
The detail of such values are presented in the Supplementary Information.

\begin{figure}[t]
	\begin{center}
	\includegraphics[width=\linewidth]{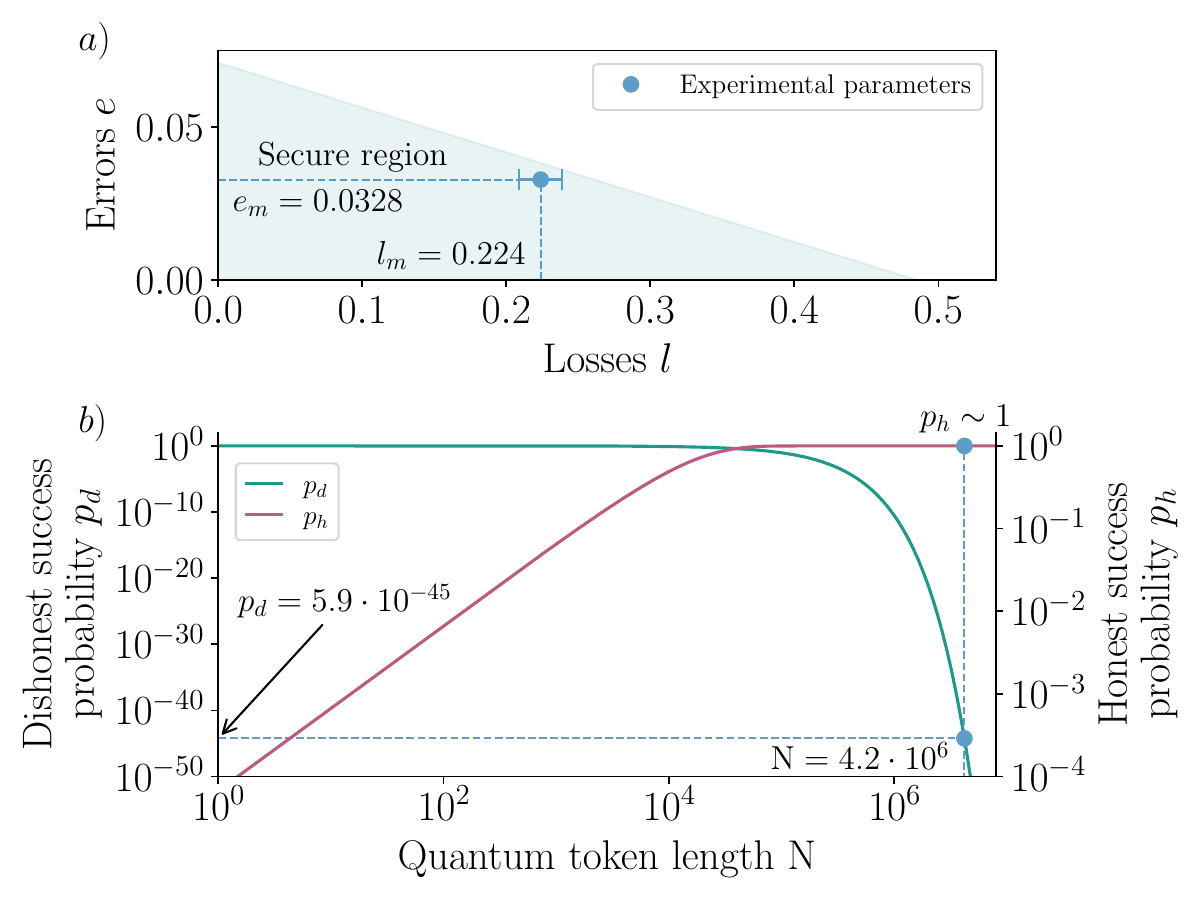}
		\caption{\textbf{Security for experimental quantum cryptograms.} \textbf{a)} The semidefinite programming framework extracts a secure region of operation (turquoise) as a function of errors and losses. Our measured experimental performance ($e_m=0.0328 \pm 0.0001; l_m=0.2239 \pm 0.015$) is indicated by the blue dot, and lies within the secure region. Error bars propagate poisson errors on coincidence counts. 
        \textbf{b)} The dishonest success probability $p_d$ (green, upper bound) and honest success probability $p_h$ (red, lower bound) are displayed as a function of the number of quantum states $N$ required to verify one bit of the cryptogram. These are derived using a Chernoff bound argument (see Supplementary Information)~\cite{DP09}. 
        As an example, an experimental token containing  $\lambda= N=4.2\cdot 10^{6}$ quantum states (vertical blue dashed line) achieves an honest success probability very close to $p_h\sim 1$ and a dishonest success probability $p_d = 5.9\cdot 10^{-45}$.}
		\label{fig:results}
	\end{center}
\end{figure}

With a maximum measured error rate of $e_m=3.28\:\pm 0.01\%$ (D/A) and losses of $l_m=22.40\:\pm 1.50\%$, we lie within the calculated secure region as depicted in \figref{fig:results}.a.
In fact, according to our semidefinite programs (see Methods), a cheating party will introduce errors larger than $e_d=3.79\:\pm 0.22\%$ when double-spending with the same amount of claimed losses $l = l_m$. With $e_m < e$ and $l_m < l$ by two standard deviations, we therefore demonstrate that a \ac{ttp} can allow for honest experimental imperfections while ensuring protection against malicious parties. 

The i.t.-secure implementation of our protocol depends only on statistical fluctuations arising from the finite number of generated quantum states: a malicious party may indeed successfully cheat by introducing fewer losses or errors than the expected asymptotic values displayed in \figref{fig:results}.a. 

We use the Chernoff bounds from \figref{fig:results}.b to estimate the \textit{dishonest success probability} $p_d$ associated with the number $N$ of quantum states required to verify one bit of the cryptogram. 
We also determine the probability $p_h$ that the protocol does not abort when followed honestly, which tends to $p_h\sim 1$ as $N$ is increased.

\medskip

\noindent\textbf{Discussion.} We propose and demonstrate a form of quantum payment that guarantees the one-time nature of purchases with \ac{itsec}. 
By increasing the length of the quantum token, the cheating probability becomes arbitrarily low in the presence of experimental imperfections such as noise and losses. The implementation does not require any challenging technology on the Client's side, besides single-photon detection.

While typical contactless payment delays are of the order of seconds, our quantum communication and verification provide \ac{itsec} within a few tens of minutes. These limitations are, however, only technological: quantum communication rates can be improved by using brighter quantum sources, while the verification delay originates from the correction of time-tagging drifts between the two buildings (see Methods). Indeed, brighter sources of entangled photon pairs have already been demonstrated, which could decrease the quantum token transmission time to under a second \cite{Ursin2021gbitsource}.

We finally note that practical digital payment schemes must allow for rejected payments without compromising the Client's sensitive data. 
In our scheme, the adversary can compromise the payment token $\ket{P}$ sent over the quantum channel, the cryptogram sent over the classical channel or the Client's choice of $M_i$.

In the first two cases, quantum mechanics will ensure that the the \ac{ttp} recognizes the malformed cryptogram and rejects the payment with arbitrarily high probability.
The transaction may than be restarted.
However, an \acp{itsec} \ac{hmac} must not re-use the key $C$ (see Methods), which is why we propose the use of $n$-time-secure \ac{hmac} to overcome this obstacle.
This allows re-using $C$ as an input for the following payments, which imposes a finite, arbitrary bound on the number of purchases~\cite{Fak.1979,rosenbaum1993lower}.
We can amend our protocol such that the number of purchases is not bounded by the
\ac{hmac} function, by growing C during the payment process: when the Client receives a new
quantum token $\ket{P}$, we append additional quantum states for QKD, and use the cryptogram  $\kappa$ for authentication.
To protect against the third case, it must be ensured that the Client's choice of $M_i$ is independent of any external bias. This can for example be guaranteed if a secure database of Merchants is initially distributed along with $C$ and the Client chooses freely without any prior communication with the Merchant.
Alternatively, the Merchant may send their ID to the Client, who uses the local database as a 2$^{nd}$ factor authentication.

Our protocol's relaxed implementation requirements with respect to previous proposals, together with its error-tolerance, facilitate its deployment in mid-term quantum networks. Classical networks host applications beyond mere communication tasks. Similarly, a future quantum internet will necessitate the maturation of various quantum primitives and applications beyond QKD \cite{Gyo:cacm22,WEH:Sci18}. Our scheme advances the field of quantum payment schemes towards mid-term practical relevancy.

\bigskip

\large
\noindent\textbf{Methods}
\normalsize

\noindent\textbf{Cryptogram.} A cryptogram is a cryptographic function that secures tokenized payments (e.g. online, contactless, and in-app-payments) against double-spending~\cite{corella2014interpreting,emvstandard}. 
The actual cryptographic mechanism varies per payment network, but a typical procedure is \textit{challenge-response}.
Here, the Client is not only in possession of a payment token, but also shares a secret key with the \ac{ttp}~\cite{Danushka.2017}.
During the payment, the Merchant generates a pseudo-random value (called a \textit{nonce}), and sends it to the client who encrypts it with this key (typically, symmetric encryption with $\ge$ 128 bit key strength is used).
The resulting \textit{cryptogram} is sent alongside the payment metadata (e.g. merchant ID, amount, etc.) to the Merchant, who forwards it to the \ac{ttp}.
As the \ac{ttp} is in possession of the key, they are able to decrypt and prove the correctness of the nonce for the given payment at the Merchant.
Spending the token for another transaction is impossible under the assumptions of computationally secure encryption.

\medskip

\noindent\textbf{\acp{itsec} \ac{hmac}.}
A \textit{\acf{hmac}} is a function $f(H,k,m) \mapsto y$.
Based on a pseudo-randomized function $H$ -- typically a hash function --, it takes a secret key $k$ and message $m$ as inputs, and outputs some authentication tag.
A hash function is defined as a function that maps a set of arbitrary length to a finite set $H:\{0,1\}^*\mapsto\{0,1\}^n; n\in\mathbb{N}$.
Hence,  hash functions are non-injective by definition, and thus collisions, such that $f(H,k,m) = f(H,k,m'); m \neq m' $ can occur (given that $k$ remains secret).
In an \acp{itsec} \ac{hmac}, the probability of such a collision is bound to $1/\sqrt{\abs{k}}; k = \{0,1\}^l$, where $l\in\mathbb{N}$ is some security parameter.
This is similar to the probability of finding the decryption key for a given one-time pad.
Different such schemes exist, in which a key $k$ can either only be used once~\cite{gilbert1974codes,Wegman.1981,YYC:NSR22},  a finite amount of times~\cite{Fak.1979,rosenbaum1993lower}, or outputted tag length is variable~\cite{Wegman.1981, Ghosh.2021}.

\medskip

\noindent\textbf{Semidefinite programming.} Our quantum-cryptographic security proof involves optimizing over semidefinite positive objects to find an adversary’s optimal cheating strategy. Semidefinite programming provides a suitable framework for this, as it allows to optimize over semidefinite positive variables, given linear constraints \cite{VB:SIAM96}. Most of the time, these variables are density matrices, measurement operators, or more general completely-positive trace-preserving maps \cite{W:LN11}. Semidefinite programs present an elegant dual structure, which associates a dual maximization problem to each primal minimization problem. The optimal value of the primal problem then upper bounds the optimal value of the dual problem, allowing to prove tight bounds on the adversarial cheating probability (see \cite{BDG:PRA19} for instance).

\medskip

\noindent\textbf{Optimal cheating strategy.}
Using semidefinite programs, we search for the optimal completely-positive trace-preserving quantum map which minimizes the introduction of noise and losses for an adversary attempting to double-spend the cryptogram. The security analysis takes into account multiphoton emission, and assumes the absence of coherence between photon number states. The latter is justified by the fact that \ac{spdc} produces states of the form $\sum_{n=0}^{\infty}\sqrt{c_n}\ket{n}_1\ket{n}_2$ in the $\{\ket{n}\}$ photon number basis \cite{BR:NJP10}, which leaves the individual subsystems in states of the form $\sum_{n=0}^{\infty}c_n\ket{n}\bra{n}$. The resulting cheating strategy is fairly intuitive when considering two extreme cases: when the tolerated error rate is zero, the malicious party splits the quantum token into two equal parts, and measures each half in a different basis. This leads to two tokens that are committed to different merchants with zero error, but with $50\%$ losses on each. On the other hand, when the tolerated losses are zero, the malicious party measures all states in a basis that is rotated by $22.5^\circ$ with respect to the H/V basis. Such a measurement will identify the correct encoded bit with a probability of $\sim85.4\%$. The actual optimal cheating strategy corresponding to our experimental parameters is a combination of these two extreme strategies.

\medskip

\noindent\textbf{State generation.}
An \ac{spdc} process in a periodically-poled KTP crystal is pumped with a continuous-wave \SI{515}{\nano\meter} laser, yielding a pair of polarization-entangled and color-entangled photons. One photon is emitted at around \SI{1500}{\nano\meter}, while its orthogonally polarized counterpart is emitted around \SI{785}{\nano\meter}. Experimental demonstrations using a similar entanglement design were demonstrated in \cite{LKH:SciRep17, RKF:npj21}. Since the spectral bandwidths of the two \ac{spdc} processes are not equal, a tunable EXFO bandpass filter is inserted into the \SI{1500}{\nano\meter} arm to equalize them and enhance the entanglement visibility. In order to render the two photons temporally indistinguishable, an unpoled KTP
crystal of half the length of the ppKTP crystal, with axes rotated by $90^\circ$ with respect to the ppKTP axes, is inserted.

\medskip

\noindent\textbf{Single-photon detection.} After the optical fiber link, the \SI{1500}{\nano\meter} photons are detected with PhotonSpot superconducting nanowire single-photon detectors, with efficiencies around $93\%$ (see Supplementary Information for detail), while the \SI{785}{\nano\meter} photons are locally detected in the \ac{ttp}'s laboratory using Roithner avalanche single-photon detectors, with efficiencies around $50\%$. A set of paddles, inserted before the polarization measurement, are used to compensate for polarization drifts over the fiber link.

\medskip

\noindent\textbf{Data post-processing.} The \ac{ttp}'s and Client's single-photon detectors are connected to two different \ac{ttm}. In order to recover coincidences between the two buildings, careful synchronisation of the two modules is required: first, the internal clocks of the respective modules bear an offset with respect
to one another, due to the photon travel time through the optical fiber link. Second, the cycles of the internal clocks of the two \acp{ttm} drift with slightly different rates, resulting in an offset drift over time. Finally, there is an electronic delay due to different detector response times, and the \acp{ttm} only record time tags relative to the time they were activated. All these factors were corrected with our post-processing code. 

\medskip

\begin{figure}[t]
    \begin{center}
        \includegraphics[width=1.0\linewidth]{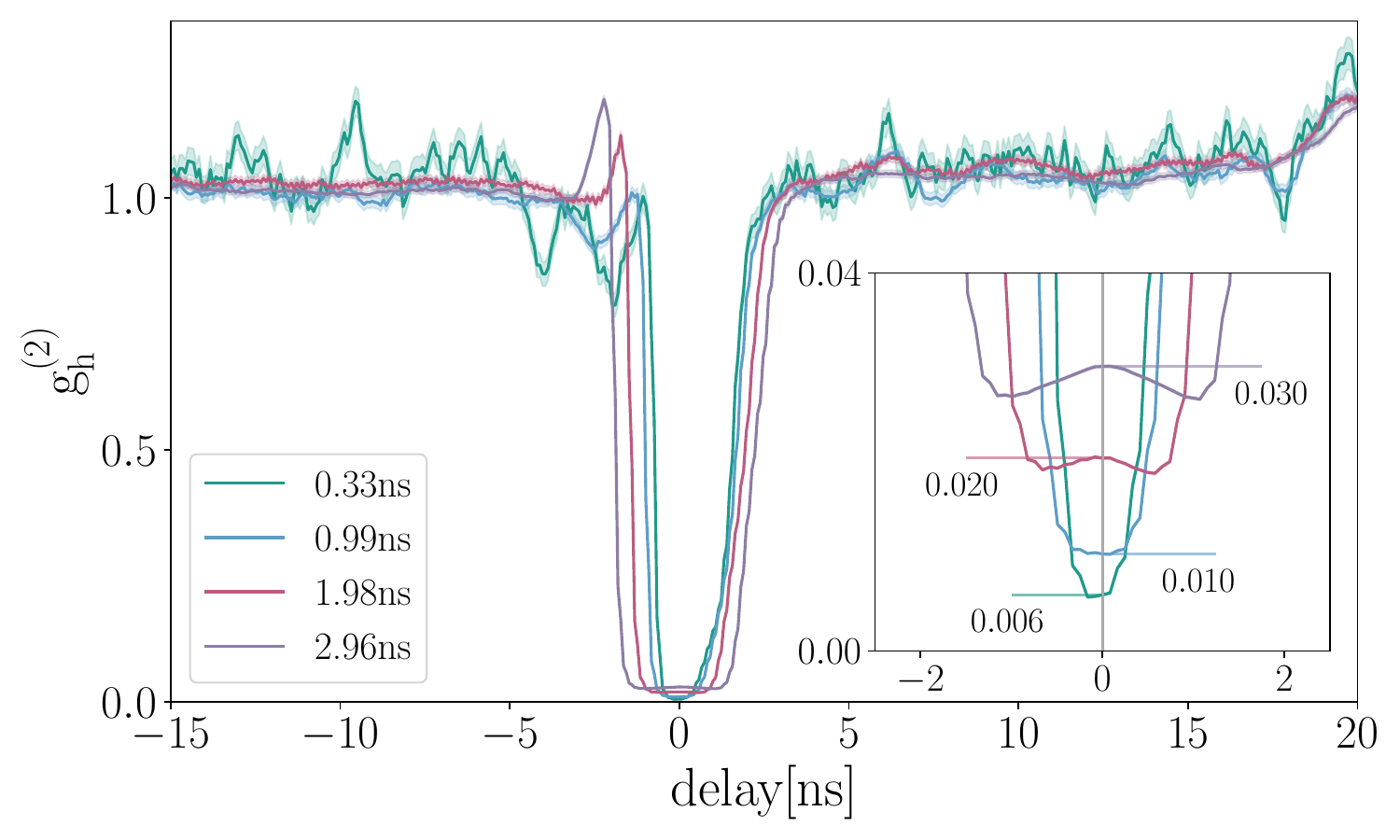}
        \caption{\textbf{Heralded second order correlation function.} Data was acquired for \SI{60}{\minute} at a pump power of \SI{35}{\milli\watt}. Coincidences were calculated using four different time windows: \SI{0.33}{\nano\second} (green), \SI{0.99}{\nano\second} (blue), \SI{1.98}{\nano\second} (red), \SI{2.96}{\nano\second} (violet). From this measurement, we determine $g^{(2)}_h(0)=\num{0.03010\pm0.00014}$ for the coincidence window used in the implementation of the protocol. Shaded areas represent error propagated uncertainties due to poissonian photon statistics.}
        \label{fig:g2}
    \end{center}
\end{figure}
\noindent\textbf{Heralded second order correlation function measurement.}
    
    To measure the heralded second order correlation function $g^{(2)}_h(\tau)$, the \SI{1500}{\nano\meter} (telecom) photons created by our \ac{spdc} source are sent directly to an InGaAs detector (idler detector labelled D\textsubscript{i}), while the \SI{785}{\nano\meter} photons are routed to a $50/50$ fiber beamsplitter, with both outputs connected to one detector each (labelled $D\textsubscript{1}$ and $D\textsubscript{2}$). $g^{(2)}_h(\tau)$ can be written as \cite{signorini2021}  
    
    \begin{equation}
        g^{(2)}_h(\tau) = \frac{N_i \cdot N_{i12}(\tau)}{N_{i1}(0) \cdot N_{i2}(\tau)},
        \label{equ:g2}
    \end{equation}
    
    \noindent where $N_i$ is the total number of events detected in the telecom detector during the measurement integration time; $N_{i1}(0)$ are the 2-fold coincidence events between the telecom detector and D1 at 0 delay; $N_{i2}(\tau)$ are the 2-fold coincidence events between the telecom detector and D2 at delay $\tau$; and $N_{i12}(\tau)$ are the 3-fold coincidences between all 3 detectors with at delay $\tau=0$ between the telcom detector and D1, and delay $\tau$ to D2. Pumping the \ac{spdc} source with \SI{35}{\milli\watt}, data was acquired for about \SI{60}{\minute}. $g^{(2)}_h(\tau)$, with coincidence time windows of \SI{0.33}{\nano\second}, \SI{0.99}{\nano\second}, \SI{1.98}{\nano\second}, \SI{2.96}{\nano\second} is shown in \figref{fig:g2}. A source dominated by single-photons has a $g^{(2)}_h(0)<0.5$, with $g^{(2)}_h(0)=0$ for a true single-photon source. From our measurements with a coincidence window of \SI{2.96}{\nano\second}, which is close to the combined jitter of the SPADs and coincidence logic and therefore the most meaningful value, we determined $g^{(2)}_h(0)=\num{0.03010\pm0.00014}$.

\medskip
\large
\noindent\textbf{Author contributions}
\normalsize

\noindent 
M.B. and P.W. conceived the project. 
J.K., E.S., T.G. and M.B. derived the security analysis. 
P.S., J.K., E.S., M-C.R., A.T. and M.B. performed the experiment and analysed the experimental data. 
P.S., J.K. and M-C.R. wrote the code required to run the experiment and process the experimental data. 
T.G. researched and explained the relevant classical digital payment schemes. 
E.S, M.B. and T.G. designed the final protocol.
P.S., J.K., E.S., T.G. and M.B. wrote the manuscript, with inputs from M-C.R., A.T. and P.W.

\medskip

\large
\noindent\textbf{Competing interests}
\normalsize

\noindent P. W., M. B., T. G., E. S. and P. S. are employees of the University of Vienna, which has applied for a patent (EP 23168897.9) for the use of a quantum payment token scheme with P. W., M. B., T. G., E. S. and P. S. listed as inventors. The remaining authors declare no competing interests.

\medskip

\large
\noindent\textbf{Data availability}
\normalsize

\noindent 
The data generated in this study have been deposited in the Zenodo database under the accession code \href{https://doi.org/10.5281/zenodo.7979319}{doi.org/10.5281/zenodo.7979319}.

\medskip 

\large
\noindent\textbf{Code availability}
\normalsize

\noindent The code used in this study has been deposited in the Zenodo database under the accession code \href{https://doi.org/10.5281/zenodo.8020667}{doi.org/10.5281/zenodo.8020667}.

\medskip 

\large
\noindent\textbf{Acknowledgements}
\normalsize

\noindent P.S., J.K., E.S., M-C.R., T.G., M.B. and P.W. acknowledge funding from the European Union's Horizon Europe research and innovation program under Grant Agreement No. 101114043 (QSNP) and No. 899368 (EPIQUS), along with the Quantum Flagship under grant No. 820474 (UNIQORN). The authors also acknowledge support from the Austrian Science Fund FWF through [F7113] (BeyondC), and [FG5] (Research Group 5); by the AFOSR via FA9550-21- 1-0355 (QTRUST); from the Austrian Federal Ministry for Digital and Economic Affairs, the National Foundation for Research, Technology and Development and the Christian Doppler Research Association. A.T. acknowledges support from the European Union’s Horizon 2020 research and innovation programme under the Marie Skłodowska-Curie grant agreement no. 801110 and the Austrian Federal Ministry of Education, Science and Research (BMBWF). For the purpose of open access, the authors have applied a CC BY public copyright licence to any Author Accepted Manuscript version arising from this submission. P.S. thanks Teodor Strömberg, and the authors thank Fofy Setaki, for fruitful discussions. 

\def\bibsection{\section*{\refname}} 
\bibliography{references}

\begin{thebibliography}{54}%
\makeatletter
\providecommand \@ifxundefined [1]{%
 \@ifx{#1\undefined}
}%
\providecommand \@ifnum [1]{%
 \ifnum #1\expandafter \@firstoftwo
 \else \expandafter \@secondoftwo
 \fi
}%
\providecommand \@ifx [1]{%
 \ifx #1\expandafter \@firstoftwo
 \else \expandafter \@secondoftwo
 \fi
}%
\providecommand \natexlab [1]{#1}%
\providecommand \enquote  [1]{``#1''}%
\providecommand \bibnamefont  [1]{#1}%
\providecommand \bibfnamefont [1]{#1}%
\providecommand \citenamefont [1]{#1}%
\providecommand \href@noop [0]{\@secondoftwo}%
\providecommand \href [0]{\begingroup \@sanitize@url \@href}%
\providecommand \@href[1]{\@@startlink{#1}\@@href}%
\providecommand \@@href[1]{\endgroup#1\@@endlink}%
\providecommand \@sanitize@url [0]{\catcode `\\12\catcode `\$12\catcode `\&12\catcode `\#12\catcode `\^12\catcode `\_12\catcode `\%12\relax}%
\providecommand \@@startlink[1]{}%
\providecommand \@@endlink[0]{}%
\providecommand \url  [0]{\begingroup\@sanitize@url \@url }%
\providecommand \@url [1]{\endgroup\@href {#1}{\urlprefix }}%
\providecommand \urlprefix  [0]{URL }%
\providecommand \Eprint [0]{\href }%
\providecommand \doibase [0]{https://doi.org/}%
\providecommand \selectlanguage [0]{\@gobble}%
\providecommand \bibinfo  [0]{\@secondoftwo}%
\providecommand \bibfield  [0]{\@secondoftwo}%
\providecommand \translation [1]{[#1]}%
\providecommand \BibitemOpen [0]{}%
\providecommand \bibitemStop [0]{}%
\providecommand \bibitemNoStop [0]{.\EOS\space}%
\providecommand \EOS [0]{\spacefactor3000\relax}%
\providecommand \BibitemShut  [1]{\csname bibitem#1\endcsname}%
\let\auto@bib@innerbib\@empty
\bibitem [{\citenamefont {Gouzien}\ and\ \citenamefont {Sangouard}(2021)}]{GS:PRL21}%
  \BibitemOpen
  \bibfield  {author} {\bibinfo {author} {\bibfnamefont {E.}~\bibnamefont {Gouzien}}\ and\ \bibinfo {author} {\bibfnamefont {N.}~\bibnamefont {Sangouard}},\ }\bibfield  {title} {\bibinfo {title} {Factoring 2048-bit rsa integers in 177 days with 13 436 qubits and a multimode memory},\ }\href {https://doi.org/10.1103/PhysRevLett.127.140503} {\bibfield  {journal} {\bibinfo  {journal} {Phys. Rev. Lett.}\ }\textbf {\bibinfo {volume} {127}},\ \bibinfo {pages} {140503} (\bibinfo {year} {2021})},\ \Eprint {https://arxiv.org/abs/2103.06159} {arXiv:2103.06159 [quant-ph]} \BibitemShut {NoStop}%
\bibitem [{\citenamefont {Martín-López}\ \emph {et~al.}(2012)\citenamefont {Martín-López}, \citenamefont {Laing}, \citenamefont {Lawson}, \citenamefont {Alvarez}, \citenamefont {Zhou},\ and\ \citenamefont {O'Brien}}]{MLL:NatPhot12}%
  \BibitemOpen
  \bibfield  {author} {\bibinfo {author} {\bibfnamefont {E.}~\bibnamefont {Martín-López}}, \bibinfo {author} {\bibfnamefont {A.}~\bibnamefont {Laing}}, \bibinfo {author} {\bibfnamefont {T.}~\bibnamefont {Lawson}}, \bibinfo {author} {\bibfnamefont {R.}~\bibnamefont {Alvarez}}, \bibinfo {author} {\bibfnamefont {X.-Q.}\ \bibnamefont {Zhou}},\ and\ \bibinfo {author} {\bibfnamefont {J.~L.}\ \bibnamefont {O'Brien}},\ }\bibfield  {title} {\bibinfo {title} {Experimental realization of shor's quantum factoring algorithm using qubit recycling},\ }\href {https://doi.org/https://www.nature.com/articles/nphoton.2012.259} {\bibfield  {journal} {\bibinfo  {journal} {Nat. Photon.}\ }\textbf {\bibinfo {volume} {6}},\ \bibinfo {pages} {773} (\bibinfo {year} {2012})},\ \Eprint {https://arxiv.org/abs/1111.4147} {arXiv:1111.4147 [quant-ph]} \BibitemShut {NoStop}%
\bibitem [{\citenamefont {Shor}(1997)}]{S:SIAM97}%
  \BibitemOpen
  \bibfield  {author} {\bibinfo {author} {\bibfnamefont {P.~W.}\ \bibnamefont {Shor}},\ }\bibfield  {title} {\bibinfo {title} {Polynomial-time algorithms for prime factorization and discrete logarithms on a quantum computer},\ }\href {https://doi.org/10.1137/S0097539795293172} {\bibfield  {journal} {\bibinfo  {journal} {SIAM Journal on Computing}\ }\textbf {\bibinfo {volume} {26}},\ \bibinfo {pages} {1484} (\bibinfo {year} {1997})}\BibitemShut {NoStop}%
\bibitem [{\citenamefont {Beullens}(2022)}]{attack-rainbow}%
  \BibitemOpen
  \bibfield  {author} {\bibinfo {author} {\bibfnamefont {W.}~\bibnamefont {Beullens}},\ }\href {https://eprint.iacr.org/2022/214} {\bibinfo {title} {Breaking rainbow takes a weekend on a laptop}},\ \bibinfo {howpublished} {Cryptology ePrint Archive, Paper 2022/214} (\bibinfo {year} {2022}),\ \bibinfo {note} {\url{https://eprint.iacr.org/2022/214}}\BibitemShut {NoStop}%
\bibitem [{\citenamefont {Castryck}\ and\ \citenamefont {Decru}(2022)}]{attack-sidh}%
  \BibitemOpen
  \bibfield  {author} {\bibinfo {author} {\bibfnamefont {W.}~\bibnamefont {Castryck}}\ and\ \bibinfo {author} {\bibfnamefont {T.}~\bibnamefont {Decru}},\ }\href {https://eprint.iacr.org/2022/975} {\bibinfo {title} {An efficient key recovery attack on sidh (preliminary version)}},\ \bibinfo {howpublished} {Cryptology ePrint Archive, Paper 2022/975} (\bibinfo {year} {2022}),\ \bibinfo {note} {\url{https://eprint.iacr.org/2022/975}}\BibitemShut {NoStop}%
\bibitem [{\citenamefont {Perlner}\ \emph {et~al.}(2022)\citenamefont {Perlner}, \citenamefont {Kelsey},\ and\ \citenamefont {Cooper}}]{attack-sphincs}%
  \BibitemOpen
  \bibfield  {author} {\bibinfo {author} {\bibfnamefont {R.}~\bibnamefont {Perlner}}, \bibinfo {author} {\bibfnamefont {J.}~\bibnamefont {Kelsey}},\ and\ \bibinfo {author} {\bibfnamefont {D.}~\bibnamefont {Cooper}},\ }\href {https://eprint.iacr.org/2022/1061} {\bibinfo {title} {Breaking category five sphincs+ with sha-256}},\ \bibinfo {howpublished} {Cryptology ePrint Archive, Paper 2022/1061} (\bibinfo {year} {2022}),\ \bibinfo {note} {\url{https://eprint.iacr.org/2022/1061}}\BibitemShut {NoStop}%
\bibitem [{\citenamefont {Bennett}\ and\ \citenamefont {Brassard}(1984)}]{BB84}%
  \BibitemOpen
  \bibfield  {author} {\bibinfo {author} {\bibfnamefont {C.~H.}\ \bibnamefont {Bennett}}\ and\ \bibinfo {author} {\bibfnamefont {G.}~\bibnamefont {Brassard}},\ }\bibfield  {title} {\bibinfo {title} {Quantum cryptography: Public key distribution and coin tossing},\ }\href {https://researcher.watson.ibm.com/researcher/files/us-bennetc/BB84highest.pdf} {\bibfield  {journal} {\bibinfo  {journal} {Proc. IEEE International Conference on Computers, Systems and Signal Processing}\ }\textbf {\bibinfo {volume} {1}},\ \bibinfo {pages} {175} (\bibinfo {year} {1984})},\ \Eprint {https://arxiv.org/abs/2003.06557} {arXiv:2003.06557 [quant-ph]} \BibitemShut {NoStop}%
\bibitem [{\citenamefont {Xu}\ \emph {et~al.}(2020)\citenamefont {Xu}, \citenamefont {Ma}, \citenamefont {Zhang}, \citenamefont {Lo},\ and\ \citenamefont {Pan}}]{Pan:RevMod20}%
  \BibitemOpen
  \bibfield  {author} {\bibinfo {author} {\bibfnamefont {F.}~\bibnamefont {Xu}}, \bibinfo {author} {\bibfnamefont {X.}~\bibnamefont {Ma}}, \bibinfo {author} {\bibfnamefont {Q.}~\bibnamefont {Zhang}}, \bibinfo {author} {\bibfnamefont {H.-K.}\ \bibnamefont {Lo}},\ and\ \bibinfo {author} {\bibfnamefont {J.-W.}\ \bibnamefont {Pan}},\ }\bibfield  {title} {\bibinfo {title} {Secure quantum key distribution with realistic devices},\ }\href {https://doi.org/10.1103/RevModPhys.92.025002} {\bibfield  {journal} {\bibinfo  {journal} {Rev. Mod. Phys.}\ }\textbf {\bibinfo {volume} {92}},\ \bibinfo {pages} {025002} (\bibinfo {year} {2020})},\ \Eprint {https://arxiv.org/abs/1903.09051} {arXiv:1903.09051 [quant-ph]} \BibitemShut {NoStop}%
\bibitem [{\citenamefont {Wehner}\ \emph {et~al.}(2018)\citenamefont {Wehner}, \citenamefont {Elkouss},\ and\ \citenamefont {Hanson}}]{WEH:Sci18}%
  \BibitemOpen
  \bibfield  {author} {\bibinfo {author} {\bibfnamefont {S.}~\bibnamefont {Wehner}}, \bibinfo {author} {\bibfnamefont {D.}~\bibnamefont {Elkouss}},\ and\ \bibinfo {author} {\bibfnamefont {R.}~\bibnamefont {Hanson}},\ }\bibfield  {title} {\bibinfo {title} {Quantum internet: A vision for the road ahead},\ }\bibfield  {journal} {\bibinfo  {journal} {Science}\ }\textbf {\bibinfo {volume} {362}},\ \href {https://doi.org/10.1126/science.aam9288} {10.1126/science.aam9288} (\bibinfo {year} {2018}),\ \Eprint {https://arxiv.org/abs/https://www.science.org/doi/pdf/10.1126/science.aam9288} {https://www.science.org/doi/pdf/10.1126/science.aam9288} \BibitemShut {NoStop}%
\bibitem [{\citenamefont {Wang}\ \emph {et~al.}(2022)\citenamefont {Wang}, \citenamefont {Yin}, \citenamefont {He}, \citenamefont {Chen}, \citenamefont {Wang}, \citenamefont {Ye}, \citenamefont {Zhou}, \citenamefont {Fan-Yuan}, \citenamefont {Wang}, \citenamefont {Wei~Chen}, \citenamefont {Morozov}, \citenamefont {Divochiy}, \citenamefont {Zhou}, \citenamefont {Guo},\ and\ \citenamefont {Han}}]{WYH:NatPhot22}%
  \BibitemOpen
  \bibfield  {author} {\bibinfo {author} {\bibfnamefont {S.}~\bibnamefont {Wang}}, \bibinfo {author} {\bibfnamefont {Z.-Q.}\ \bibnamefont {Yin}}, \bibinfo {author} {\bibfnamefont {D.-Y.}\ \bibnamefont {He}}, \bibinfo {author} {\bibfnamefont {W.}~\bibnamefont {Chen}}, \bibinfo {author} {\bibfnamefont {R.-Q.}\ \bibnamefont {Wang}}, \bibinfo {author} {\bibfnamefont {P.}~\bibnamefont {Ye}}, \bibinfo {author} {\bibfnamefont {Y.}~\bibnamefont {Zhou}}, \bibinfo {author} {\bibfnamefont {G.-J.}\ \bibnamefont {Fan-Yuan}}, \bibinfo {author} {\bibfnamefont {F.-X.}\ \bibnamefont {Wang}}, \bibinfo {author} {\bibfnamefont {Y.-G.~Z.}\ \bibnamefont {Wei~Chen}}, \bibinfo {author} {\bibfnamefont {P.~V.}\ \bibnamefont {Morozov}}, \bibinfo {author} {\bibfnamefont {A.~V.}\ \bibnamefont {Divochiy}}, \bibinfo {author} {\bibfnamefont {Z.}~\bibnamefont {Zhou}}, \bibinfo {author} {\bibfnamefont {G.-C.}\ \bibnamefont {Guo}},\ and\ \bibinfo {author} {\bibfnamefont {Z.-F.}\ \bibnamefont {Han}},\ }\bibfield  {title} {\bibinfo {title}
  {Twin-field quantum key distribution over 830-km fibre},\ }\href {https://doi.org/10.1038/s41566-021-00928-2} {\bibfield  {journal} {\bibinfo  {journal} {Nat. Photon.}\ }\textbf {\bibinfo {volume} {16}},\ \bibinfo {pages} {154} (\bibinfo {year} {2022})}\BibitemShut {NoStop}%
\bibitem [{\citenamefont {Boaron}\ \emph {et~al.}(2018)\citenamefont {Boaron}, \citenamefont {Boso}, \citenamefont {Rusca}, \citenamefont {Vulliez}, \citenamefont {Autebert}, \citenamefont {Caloz}, \citenamefont {Perrenoud}, \citenamefont {Gras}, \citenamefont {Bussi\`eres}, \citenamefont {Li}, \citenamefont {Nolan}, \citenamefont {Martin},\ and\ \citenamefont {Zbinden}}]{BBR:PRL18}%
  \BibitemOpen
  \bibfield  {author} {\bibinfo {author} {\bibfnamefont {A.}~\bibnamefont {Boaron}}, \bibinfo {author} {\bibfnamefont {G.}~\bibnamefont {Boso}}, \bibinfo {author} {\bibfnamefont {D.}~\bibnamefont {Rusca}}, \bibinfo {author} {\bibfnamefont {C.}~\bibnamefont {Vulliez}}, \bibinfo {author} {\bibfnamefont {C.}~\bibnamefont {Autebert}}, \bibinfo {author} {\bibfnamefont {M.}~\bibnamefont {Caloz}}, \bibinfo {author} {\bibfnamefont {M.}~\bibnamefont {Perrenoud}}, \bibinfo {author} {\bibfnamefont {G.}~\bibnamefont {Gras}}, \bibinfo {author} {\bibfnamefont {F.}~\bibnamefont {Bussi\`eres}}, \bibinfo {author} {\bibfnamefont {M.-J.}\ \bibnamefont {Li}}, \bibinfo {author} {\bibfnamefont {D.}~\bibnamefont {Nolan}}, \bibinfo {author} {\bibfnamefont {A.}~\bibnamefont {Martin}},\ and\ \bibinfo {author} {\bibfnamefont {H.}~\bibnamefont {Zbinden}},\ }\bibfield  {title} {\bibinfo {title} {Secure quantum key distribution over 421 km of optical fiber},\ }\href {https://doi.org/10.1103/PhysRevLett.121.190502} {\bibfield  {journal}
  {\bibinfo  {journal} {Phys. Rev. Lett.}\ }\textbf {\bibinfo {volume} {121}},\ \bibinfo {pages} {190502} (\bibinfo {year} {2018})},\ \Eprint {https://arxiv.org/abs/1807.03222} {arXiv:1807.03222 [quant-ph]} \BibitemShut {NoStop}%
\bibitem [{\citenamefont {Yin}\ \emph {et~al.}(2020)\citenamefont {Yin}, \citenamefont {Li},\ and\ \citenamefont {Liao}}]{Pan:Nature20}%
  \BibitemOpen
  \bibfield  {author} {\bibinfo {author} {\bibfnamefont {J.}~\bibnamefont {Yin}}, \bibinfo {author} {\bibfnamefont {Y.}~\bibnamefont {Li}},\ and\ \bibinfo {author} {\bibfnamefont {S.~e.~a.}\ \bibnamefont {Liao}},\ }\bibfield  {title} {\bibinfo {title} {Entanglement-based secure quantum cryptography over 1,120 kilometres},\ }\href {https://doi.org/10.1038/s41586-020-2401-y} {\bibfield  {journal} {\bibinfo  {journal} {Nature}\ }\textbf {\bibinfo {volume} {582}},\ \bibinfo {pages} {501} (\bibinfo {year} {2020})}\BibitemShut {NoStop}%
\bibitem [{\citenamefont {Bedington}\ \emph {et~al.}(2017)\citenamefont {Bedington}, \citenamefont {Arrazola},\ and\ \citenamefont {Ling}}]{BAL:npjqi17}%
  \BibitemOpen
  \bibfield  {author} {\bibinfo {author} {\bibfnamefont {R.}~\bibnamefont {Bedington}}, \bibinfo {author} {\bibfnamefont {J.-M.}\ \bibnamefont {Arrazola}},\ and\ \bibinfo {author} {\bibfnamefont {A.}~\bibnamefont {Ling}},\ }\bibfield  {title} {\bibinfo {title} {Progress in satellite quantum key distribution},\ }\href {https://doi.org/10.1038/s41534-017-0031-5} {\bibfield  {journal} {\bibinfo  {journal} {npj Quantum Inf.}\ }\textbf {\bibinfo {volume} {3}},\ \bibinfo {pages} {30} (\bibinfo {year} {2017})},\ \Eprint {https://arxiv.org/abs/1707.03613} {arXiv:1707.03613 [quant-ph]} \BibitemShut {NoStop}%
\bibitem [{\citenamefont {{PCI Security Standards Council (PCI SSC)}}()}]{PCIstandard}%
  \BibitemOpen
  \bibfield  {author} {\bibinfo {author} {\bibnamefont {{PCI Security Standards Council (PCI SSC)}}},\ }\href@noop {} {}\bibinfo {howpublished} {\url{https://www.pcisecuritystandards.org}},\ \bibinfo {note} {accessed: 2022-11-02}\BibitemShut {NoStop}%
\bibitem [{\citenamefont {Corella}\ and\ \citenamefont {Lewison}(2014)}]{corella2014interpreting}%
  \BibitemOpen
  \bibfield  {author} {\bibinfo {author} {\bibfnamefont {F.}~\bibnamefont {Corella}}\ and\ \bibinfo {author} {\bibfnamefont {K.}~\bibnamefont {Lewison}},\ }\bibfield  {title} {\bibinfo {title} {{Interpreting the EMV tokenisation specification}},\ }\href {https://pomcor.com/whitepapers/EMVTok.pdf} {\bibfield  {journal} {\bibinfo  {journal} {white paper}\ } (\bibinfo {year} {2014})}\BibitemShut {NoStop}%
\bibitem [{\citenamefont {{EMVCo\ LLC}}(2021)}]{emvstandard}%
  \BibitemOpen
  \bibfield  {author} {\bibinfo {author} {\bibnamefont {{EMVCo\ LLC}}},\ }\href {https://www.emvco.com/specifications/} {\bibinfo {title} {{EMV Payment Tokenisation Specification -- Technical Framework}}} (\bibinfo {year} {2021})\BibitemShut {NoStop}%
\bibitem [{\citenamefont {Zhang}\ \emph {et~al.}(2022)\citenamefont {Zhang}, \citenamefont {van Leent}, \citenamefont {Redeker}, \citenamefont {Garthoff}, \citenamefont {Schwonnek}, \citenamefont {Fertig}, \citenamefont {Eppelt}, \citenamefont {Rosenfeld}, \citenamefont {Scarani}, \citenamefont {Lim},\ and\ \citenamefont {Weinfurter}}]{ZLR:Nature22}%
  \BibitemOpen
  \bibfield  {author} {\bibinfo {author} {\bibfnamefont {W.}~\bibnamefont {Zhang}}, \bibinfo {author} {\bibfnamefont {T.}~\bibnamefont {van Leent}}, \bibinfo {author} {\bibfnamefont {K.}~\bibnamefont {Redeker}}, \bibinfo {author} {\bibfnamefont {R.}~\bibnamefont {Garthoff}}, \bibinfo {author} {\bibfnamefont {R.}~\bibnamefont {Schwonnek}}, \bibinfo {author} {\bibfnamefont {F.}~\bibnamefont {Fertig}}, \bibinfo {author} {\bibfnamefont {S.}~\bibnamefont {Eppelt}}, \bibinfo {author} {\bibfnamefont {W.}~\bibnamefont {Rosenfeld}}, \bibinfo {author} {\bibfnamefont {V.}~\bibnamefont {Scarani}}, \bibinfo {author} {\bibfnamefont {C.~C.-W.}\ \bibnamefont {Lim}},\ and\ \bibinfo {author} {\bibfnamefont {H.}~\bibnamefont {Weinfurter}},\ }\bibfield  {title} {\bibinfo {title} {A device-independent quantum key distribution system for distant users},\ }\href {https://doi.org/10.1038/s41586-022-04891-y} {\bibfield  {journal} {\bibinfo  {journal} {Nature}\ }\textbf {\bibinfo {volume} {607}},\ \bibinfo {pages} {687–691}
  (\bibinfo {year} {2022})},\ \Eprint {https://arxiv.org/abs/2110.00575} {arXiv:2110.00575 [quant-ph]} \BibitemShut {NoStop}%
\bibitem [{\citenamefont {Nadlinger}\ \emph {et~al.}(2022)\citenamefont {Nadlinger}, \citenamefont {Drmota}, \citenamefont {Nichol}, \citenamefont {Araneda}, \citenamefont {Main}, \citenamefont {Srinivas}, \citenamefont {Lucas}, \citenamefont {Ballance}, \citenamefont {Ivanov}, \citenamefont {Tan}, \citenamefont {Sekatski}, \citenamefont {Urbanke}, \citenamefont {Renner}, \citenamefont {Sangouard},\ and\ \citenamefont {Bancal}}]{NDN:Nature22}%
  \BibitemOpen
  \bibfield  {author} {\bibinfo {author} {\bibfnamefont {D.~P.}\ \bibnamefont {Nadlinger}}, \bibinfo {author} {\bibfnamefont {P.}~\bibnamefont {Drmota}}, \bibinfo {author} {\bibfnamefont {B.~C.}\ \bibnamefont {Nichol}}, \bibinfo {author} {\bibfnamefont {G.}~\bibnamefont {Araneda}}, \bibinfo {author} {\bibfnamefont {D.}~\bibnamefont {Main}}, \bibinfo {author} {\bibfnamefont {R.}~\bibnamefont {Srinivas}}, \bibinfo {author} {\bibfnamefont {D.~M.}\ \bibnamefont {Lucas}}, \bibinfo {author} {\bibfnamefont {C.~J.}\ \bibnamefont {Ballance}}, \bibinfo {author} {\bibfnamefont {K.}~\bibnamefont {Ivanov}}, \bibinfo {author} {\bibfnamefont {E.~Y.-Z.}\ \bibnamefont {Tan}}, \bibinfo {author} {\bibfnamefont {P.}~\bibnamefont {Sekatski}}, \bibinfo {author} {\bibfnamefont {R.~L.}\ \bibnamefont {Urbanke}}, \bibinfo {author} {\bibfnamefont {R.}~\bibnamefont {Renner}}, \bibinfo {author} {\bibfnamefont {N.}~\bibnamefont {Sangouard}},\ and\ \bibinfo {author} {\bibfnamefont {J.-D.}\ \bibnamefont {Bancal}},\ }\bibfield  {title}
  {\bibinfo {title} {Experimental quantum key distribution certified by bell's theorem},\ }\href {https://doi.org/10.1038/s41586-022-04941-5} {\bibfield  {journal} {\bibinfo  {journal} {Nature}\ }\textbf {\bibinfo {volume} {607}},\ \bibinfo {pages} {682–686} (\bibinfo {year} {2022})},\ \Eprint {https://arxiv.org/abs/2109.14600} {arXiv:2109.14600 [quant-ph]} \BibitemShut {NoStop}%
\bibitem [{\citenamefont {Liu}\ \emph {et~al.}(2022)\citenamefont {Liu}, \citenamefont {Zhang}, \citenamefont {Zhen}, \citenamefont {Li}, \citenamefont {Liu}, \citenamefont {Fan}, \citenamefont {Xu}, \citenamefont {Zhang},\ and\ \citenamefont {Pan}}]{LZZ:PRL22}%
  \BibitemOpen
  \bibfield  {author} {\bibinfo {author} {\bibfnamefont {W.-Z.}\ \bibnamefont {Liu}}, \bibinfo {author} {\bibfnamefont {Y.-Z.}\ \bibnamefont {Zhang}}, \bibinfo {author} {\bibfnamefont {Y.-Z.}\ \bibnamefont {Zhen}}, \bibinfo {author} {\bibfnamefont {M.-H.}\ \bibnamefont {Li}}, \bibinfo {author} {\bibfnamefont {Y.}~\bibnamefont {Liu}}, \bibinfo {author} {\bibfnamefont {J.}~\bibnamefont {Fan}}, \bibinfo {author} {\bibfnamefont {F.}~\bibnamefont {Xu}}, \bibinfo {author} {\bibfnamefont {Q.}~\bibnamefont {Zhang}},\ and\ \bibinfo {author} {\bibfnamefont {J.-W.}\ \bibnamefont {Pan}},\ }\bibfield  {title} {\bibinfo {title} {Toward a photonic demonstration of device-independent quantum key distribution},\ }\href {https://doi.org/10.1103/PhysRevLett.129.050502} {\bibfield  {journal} {\bibinfo  {journal} {Phys. Rev. Lett.}\ }\textbf {\bibinfo {volume} {129}},\ \bibinfo {pages} {050502} (\bibinfo {year} {2022})}\BibitemShut {NoStop}%
\bibitem [{\citenamefont {Wiesner}(1983)}]{Wie:acm83}%
  \BibitemOpen
  \bibfield  {author} {\bibinfo {author} {\bibfnamefont {S.}~\bibnamefont {Wiesner}},\ }\bibfield  {title} {\bibinfo {title} {Conjugate coding},\ }\href {https://doi.org/10.1145/1008908.1008920} {\bibfield  {journal} {\bibinfo  {journal} {ACM Sigact News}\ }\textbf {\bibinfo {volume} {15}},\ \bibinfo {pages} {78} (\bibinfo {year} {1983})}\BibitemShut {NoStop}%
\bibitem [{\citenamefont {Aaronson}\ and\ \citenamefont {Christiano}(2012)}]{AC:stoc12}%
  \BibitemOpen
  \bibfield  {author} {\bibinfo {author} {\bibfnamefont {S.}~\bibnamefont {Aaronson}}\ and\ \bibinfo {author} {\bibfnamefont {P.}~\bibnamefont {Christiano}},\ }\bibfield  {title} {\bibinfo {title} {Quantum money from hidden subspaces},\ }\href {https://doi.org/10.1145/2213977.2213983} {\bibfield  {journal} {\bibinfo  {journal} {Proceedings of the Forty-Fourth Annual ACM Symposium on Theory of Computing}\ ,\ \bibinfo {pages} {41–60}} (\bibinfo {year} {2012})},\ \Eprint {https://arxiv.org/abs/1203.4740} {arXiv:1203.4740 [quant-ph]} \BibitemShut {NoStop}%
\bibitem [{\citenamefont {Bartkiewicz}\ \emph {et~al.}(2017)\citenamefont {Bartkiewicz}, \citenamefont {\v{C}ernoch}, \citenamefont {Chimczak}, \citenamefont {Lemr}, \citenamefont {Miranowicz},\ and\ \citenamefont {Nori}}]{BC+:npjQI17}%
  \BibitemOpen
  \bibfield  {author} {\bibinfo {author} {\bibfnamefont {K.}~\bibnamefont {Bartkiewicz}}, \bibinfo {author} {\bibfnamefont {A.}~\bibnamefont {\v{C}ernoch}}, \bibinfo {author} {\bibfnamefont {G.}~\bibnamefont {Chimczak}}, \bibinfo {author} {\bibfnamefont {K.}~\bibnamefont {Lemr}}, \bibinfo {author} {\bibfnamefont {A.}~\bibnamefont {Miranowicz}},\ and\ \bibinfo {author} {\bibfnamefont {F.}~\bibnamefont {Nori}},\ }\bibfield  {title} {\bibinfo {title} {Experimental quantum forgery of quantum optical money},\ }\href {https://doi.org/10.1038/s41534-017-0010-x} {\bibfield  {journal} {\bibinfo  {journal} {npj Quantum Inf.}\ }\textbf {\bibinfo {volume} {3}},\ \bibinfo {pages} {7} (\bibinfo {year} {2017})},\ \Eprint {https://arxiv.org/abs/1604.04453} {arXiv:1604.04453 [quant-ph]} \BibitemShut {NoStop}%
\bibitem [{\citenamefont {Pastawski}\ \emph {et~al.}(2012)\citenamefont {Pastawski}, \citenamefont {Yao}, \citenamefont {Jiang}, \citenamefont {Lukin},\ and\ \citenamefont {Cirac}}]{PY+:pnas12}%
  \BibitemOpen
  \bibfield  {author} {\bibinfo {author} {\bibfnamefont {F.}~\bibnamefont {Pastawski}}, \bibinfo {author} {\bibfnamefont {N.~Y.}\ \bibnamefont {Yao}}, \bibinfo {author} {\bibfnamefont {L.}~\bibnamefont {Jiang}}, \bibinfo {author} {\bibfnamefont {M.~D.}\ \bibnamefont {Lukin}},\ and\ \bibinfo {author} {\bibfnamefont {J.~I.}\ \bibnamefont {Cirac}},\ }\bibfield  {title} {\bibinfo {title} {Unforgeable noise-tolerant quantum tokens},\ }\href {https://doi.org/10.1073/pnas.1203552109} {\bibfield  {journal} {\bibinfo  {journal} {PNAS}\ }\textbf {\bibinfo {volume} {109}},\ \bibinfo {pages} {16079} (\bibinfo {year} {2012})},\ \Eprint {https://arxiv.org/abs/1112.5456} {arXiv:1112.5456 [quant-ph]} \BibitemShut {NoStop}%
\bibitem [{\citenamefont {{Bozzio}}\ \emph {et~al.}(2018)\citenamefont {{Bozzio}}, \citenamefont {{Orieux}}, \citenamefont {{Trigo Vidarte}}, \citenamefont {{Zaquine}}, \citenamefont {{Kerenidis}},\ and\ \citenamefont {{Diamanti}}}]{BOV:npj18}%
  \BibitemOpen
  \bibfield  {author} {\bibinfo {author} {\bibfnamefont {M.}~\bibnamefont {{Bozzio}}}, \bibinfo {author} {\bibfnamefont {A.}~\bibnamefont {{Orieux}}}, \bibinfo {author} {\bibfnamefont {L.}~\bibnamefont {{Trigo Vidarte}}}, \bibinfo {author} {\bibfnamefont {I.}~\bibnamefont {{Zaquine}}}, \bibinfo {author} {\bibfnamefont {I.}~\bibnamefont {{Kerenidis}}},\ and\ \bibinfo {author} {\bibfnamefont {E.}~\bibnamefont {{Diamanti}}},\ }\bibfield  {title} {\bibinfo {title} {{Experimental investigation of practical unforgeable quantum money}},\ }\href {https://doi.org/10.1038/s41534-018-0058-2} {\bibfield  {journal} {\bibinfo  {journal} {npj Quantum Inf.}\ }\textbf {\bibinfo {volume} {4}},\ \bibinfo {eid} {5} (\bibinfo {year} {2018})},\ \Eprint {https://arxiv.org/abs/1705.01428} {arXiv:1705.01428 [quant-ph]} \BibitemShut {NoStop}%
\bibitem [{\citenamefont {Guan}\ \emph {et~al.}(2018)\citenamefont {Guan}, \citenamefont {Arrazola}, \citenamefont {Amiri}, \citenamefont {Zhang}, \citenamefont {Li}, \citenamefont {You}, \citenamefont {Wang}, \citenamefont {Zhang},\ and\ \citenamefont {Pan}}]{GAA:pra18}%
  \BibitemOpen
  \bibfield  {author} {\bibinfo {author} {\bibfnamefont {J.-Y.}\ \bibnamefont {Guan}}, \bibinfo {author} {\bibfnamefont {J.-M.}\ \bibnamefont {Arrazola}}, \bibinfo {author} {\bibfnamefont {R.}~\bibnamefont {Amiri}}, \bibinfo {author} {\bibfnamefont {W.}~\bibnamefont {Zhang}}, \bibinfo {author} {\bibfnamefont {H.}~\bibnamefont {Li}}, \bibinfo {author} {\bibfnamefont {L.}~\bibnamefont {You}}, \bibinfo {author} {\bibfnamefont {Z.}~\bibnamefont {Wang}}, \bibinfo {author} {\bibfnamefont {Q.}~\bibnamefont {Zhang}},\ and\ \bibinfo {author} {\bibfnamefont {J.-W.}\ \bibnamefont {Pan}},\ }\bibfield  {title} {\bibinfo {title} {Experimental preparation and verification of quantum money},\ }\href {https://doi.org/10.1103/PhysRevA.97.032338} {\bibfield  {journal} {\bibinfo  {journal} {Phys. Rev. A}\ }\textbf {\bibinfo {volume} {97}},\ \bibinfo {pages} {032338} (\bibinfo {year} {2018})},\ \Eprint {https://arxiv.org/abs/1709.05882} {arXiv:1709.05882 [quant-ph]} \BibitemShut {NoStop}%
\bibitem [{\citenamefont {{Bozzio}}\ \emph {et~al.}(2019)\citenamefont {{Bozzio}}, \citenamefont {{Diamanti}},\ and\ \citenamefont {{Grosshans}}}]{BDG:PRA19}%
  \BibitemOpen
  \bibfield  {author} {\bibinfo {author} {\bibfnamefont {M.}~\bibnamefont {{Bozzio}}}, \bibinfo {author} {\bibfnamefont {E.}~\bibnamefont {{Diamanti}}},\ and\ \bibinfo {author} {\bibfnamefont {F.}~\bibnamefont {{Grosshans}}},\ }\bibfield  {title} {\bibinfo {title} {{Semi-device-independent quantum money with coherent states}},\ }\href {https://doi.org/10.1103/PhysRevA.99.022336} {\bibfield  {journal} {\bibinfo  {journal} {Physical Review A: General Physics}\ }\textbf {\bibinfo {volume} {99}},\ \bibinfo {eid} {022336} (\bibinfo {year} {2019})},\ \Eprint {https://arxiv.org/abs/1812.09256} {arXiv:1812.09256 [quant-ph]} \BibitemShut {NoStop}%
\bibitem [{\citenamefont {Horodecki}\ and\ \citenamefont {Stankiewicz}(2020)}]{HS:NJP20}%
  \BibitemOpen
  \bibfield  {author} {\bibinfo {author} {\bibfnamefont {K.}~\bibnamefont {Horodecki}}\ and\ \bibinfo {author} {\bibfnamefont {M.}~\bibnamefont {Stankiewicz}},\ }\bibfield  {title} {\bibinfo {title} {Semi-device-independent quantum money},\ }\href {https://doi.org/10.1088/1367-2630/ab6872} {\bibfield  {journal} {\bibinfo  {journal} {New J. Phys.}\ }\textbf {\bibinfo {volume} {22}},\ \bibinfo {pages} {023007} (\bibinfo {year} {2020})},\ \Eprint {https://arxiv.org/abs/1811.10552} {arXiv:1811.10552 [quant-ph]} \BibitemShut {NoStop}%
\bibitem [{\citenamefont {{Ma}}\ \emph {et~al.}(2021)\citenamefont {{Ma}}, \citenamefont {{Ma}}, \citenamefont {{Zhou}}, \citenamefont {{Li}},\ and\ \citenamefont {{Guo}}}]{MMZ:natcomms21}%
  \BibitemOpen
  \bibfield  {author} {\bibinfo {author} {\bibfnamefont {Y.}~\bibnamefont {{Ma}}}, \bibinfo {author} {\bibfnamefont {Y.-Z.}\ \bibnamefont {{Ma}}}, \bibinfo {author} {\bibfnamefont {Z.-Q.}\ \bibnamefont {{Zhou}}}, \bibinfo {author} {\bibfnamefont {C.-F.}\ \bibnamefont {{Li}}},\ and\ \bibinfo {author} {\bibfnamefont {G.-C.}\ \bibnamefont {{Guo}}},\ }\bibfield  {title} {\bibinfo {title} {{One-hour coherent optical storage in an atomic frequency comb memory}},\ }\href {https://doi.org/10.1038/s41467-021-22706-y} {\bibfield  {journal} {\bibinfo  {journal} {Nat. Commun.}\ }\textbf {\bibinfo {volume} {12}},\ \bibinfo {eid} {2381} (\bibinfo {year} {2021})},\ \Eprint {https://arxiv.org/abs/2012.14605} {arXiv:2012.14605 [quant-ph]} \BibitemShut {NoStop}%
\bibitem [{\citenamefont {{Vernaz-Gris}}\ \emph {et~al.}(2018)\citenamefont {{Vernaz-Gris}}, \citenamefont {{Huang}}, \citenamefont {{Cao}}, \citenamefont {{Sheremet}},\ and\ \citenamefont {{Laurat}}}]{VHC:NC18}%
  \BibitemOpen
  \bibfield  {author} {\bibinfo {author} {\bibfnamefont {P.}~\bibnamefont {{Vernaz-Gris}}}, \bibinfo {author} {\bibfnamefont {K.}~\bibnamefont {{Huang}}}, \bibinfo {author} {\bibfnamefont {M.}~\bibnamefont {{Cao}}}, \bibinfo {author} {\bibfnamefont {A.~S.}\ \bibnamefont {{Sheremet}}},\ and\ \bibinfo {author} {\bibfnamefont {J.}~\bibnamefont {{Laurat}}},\ }\bibfield  {title} {\bibinfo {title} {{Highly-efficient quantum memory for polarization qubits in a spatially-multiplexed cold atomic ensemble}},\ }\href {https://doi.org/10.1038/s41467-017-02775-8} {\bibfield  {journal} {\bibinfo  {journal} {Nat. Commun.}\ }\textbf {\bibinfo {volume} {9}},\ \bibinfo {eid} {363} (\bibinfo {year} {2018})},\ \Eprint {https://arxiv.org/abs/1707.09372} {arXiv:1707.09372 [quant-ph]} \BibitemShut {NoStop}%
\bibitem [{\citenamefont {Heshami}\ \emph {et~al.}(2016)\citenamefont {Heshami}, \citenamefont {England}, \citenamefont {Humphreys}, \citenamefont {Bustard}, \citenamefont {Acosta}, \citenamefont {Nunn},\ and\ \citenamefont {Sussman}}]{HE+:jmo16}%
  \BibitemOpen
  \bibfield  {author} {\bibinfo {author} {\bibfnamefont {K.}~\bibnamefont {Heshami}}, \bibinfo {author} {\bibfnamefont {D.~G.}\ \bibnamefont {England}}, \bibinfo {author} {\bibfnamefont {P.~C.}\ \bibnamefont {Humphreys}}, \bibinfo {author} {\bibfnamefont {P.~J.}\ \bibnamefont {Bustard}}, \bibinfo {author} {\bibfnamefont {V.~M.}\ \bibnamefont {Acosta}}, \bibinfo {author} {\bibfnamefont {J.}~\bibnamefont {Nunn}},\ and\ \bibinfo {author} {\bibfnamefont {B.~J.}\ \bibnamefont {Sussman}},\ }\bibfield  {title} {\bibinfo {title} {Quantum memories: emerging applications and recent advances},\ }\href {https://doi.org/10.1080/09500340.2016.1148212} {\bibfield  {journal} {\bibinfo  {journal} {J. Mod. Opt.}\ }\textbf {\bibinfo {volume} {63}},\ \bibinfo {pages} {2005} (\bibinfo {year} {2016})},\ \Eprint {https://arxiv.org/abs/1511.04018} {arXiv:1511.04018 [quant-ph]} \BibitemShut {NoStop}%
\bibitem [{\citenamefont {{Kent}}\ and\ \citenamefont {{Pital{\'u}a-Garc{\'\i}a}}(2020)}]{Kent:PRA20}%
  \BibitemOpen
  \bibfield  {author} {\bibinfo {author} {\bibfnamefont {A.}~\bibnamefont {{Kent}}}\ and\ \bibinfo {author} {\bibfnamefont {D.}~\bibnamefont {{Pital{\'u}a-Garc{\'\i}a}}},\ }\bibfield  {title} {\bibinfo {title} {{Flexible quantum tokens in spacetime}},\ }\href {https://doi.org/10.1103/PhysRevA.101.022309} {\bibfield  {journal} {\bibinfo  {journal} {Physical Review A: General Physics}\ }\textbf {\bibinfo {volume} {101}},\ \bibinfo {eid} {022309} (\bibinfo {year} {2020})},\ \Eprint {https://arxiv.org/abs/1908.08143} {arXiv:1908.08143 [quant-ph]} \BibitemShut {NoStop}%
\bibitem [{\citenamefont {{Kent}}\ \emph {et~al.}(2022)\citenamefont {{Kent}}, \citenamefont {{Lowndes}}, \citenamefont {{Pital{\'u}a-Garc{\'\i}a}},\ and\ \citenamefont {{Rarity}}}]{Kent:npj22}%
  \BibitemOpen
  \bibfield  {author} {\bibinfo {author} {\bibfnamefont {A.}~\bibnamefont {{Kent}}}, \bibinfo {author} {\bibfnamefont {D.}~\bibnamefont {{Lowndes}}}, \bibinfo {author} {\bibfnamefont {D.}~\bibnamefont {{Pital{\'u}a-Garc{\'\i}a}}},\ and\ \bibinfo {author} {\bibfnamefont {J.}~\bibnamefont {{Rarity}}},\ }\bibfield  {title} {\bibinfo {title} {{Practical quantum tokens without quantum memories and experimental tests}},\ }\href {https://doi.org/10.1038/s41534-022-00524-4} {\bibfield  {journal} {\bibinfo  {journal} {npj Quantum Inf.}\ }\textbf {\bibinfo {volume} {8}},\ \bibinfo {eid} {28} (\bibinfo {year} {2022})},\ \Eprint {https://arxiv.org/abs/2104.11717} {arXiv:2104.11717 [quant-ph]} \BibitemShut {NoStop}%
\bibitem [{\citenamefont {Tippenhauer}\ \emph {et~al.}(2011)\citenamefont {Tippenhauer}, \citenamefont {P\"{o}pper}, \citenamefont {Rasmussen},\ and\ \citenamefont {Capkun}}]{spoof11}%
  \BibitemOpen
  \bibfield  {author} {\bibinfo {author} {\bibfnamefont {N.~O.}\ \bibnamefont {Tippenhauer}}, \bibinfo {author} {\bibfnamefont {C.}~\bibnamefont {P\"{o}pper}}, \bibinfo {author} {\bibfnamefont {K.~B.}\ \bibnamefont {Rasmussen}},\ and\ \bibinfo {author} {\bibfnamefont {S.}~\bibnamefont {Capkun}},\ }\bibfield  {title} {\bibinfo {title} {On the requirements for successful gps spoofing attacks},\ }\href {https://doi.org/10.1145/2046707.2046719} {\bibfield  {journal} {\bibinfo  {journal} {Proc. of the 18th ACM Conference on Computer and Communications Security}\ }\bibinfo {series} {CCS '11},\ \bibinfo {pages} {75–86} (\bibinfo {year} {2011})}\BibitemShut {NoStop}%
\bibitem [{\citenamefont {Bozzio}\ \emph {et~al.}(2021)\citenamefont {Bozzio}, \citenamefont {Cavaill\`es}, \citenamefont {Diamanti}, \citenamefont {Kent},\ and\ \citenamefont {Pital\'ua-Garc\'{\i}a}}]{BCD:PRX21}%
  \BibitemOpen
  \bibfield  {author} {\bibinfo {author} {\bibfnamefont {M.}~\bibnamefont {Bozzio}}, \bibinfo {author} {\bibfnamefont {A.}~\bibnamefont {Cavaill\`es}}, \bibinfo {author} {\bibfnamefont {E.}~\bibnamefont {Diamanti}}, \bibinfo {author} {\bibfnamefont {A.}~\bibnamefont {Kent}},\ and\ \bibinfo {author} {\bibfnamefont {D.}~\bibnamefont {Pital\'ua-Garc\'{\i}a}},\ }\bibfield  {title} {\bibinfo {title} {Multiphoton and side-channel attacks in mistrustful quantum cryptography},\ }\href {https://doi.org/10.1103/PRXQuantum.2.030338} {\bibfield  {journal} {\bibinfo  {journal} {PRX Quantum}\ }\textbf {\bibinfo {volume} {2}},\ \bibinfo {pages} {030338} (\bibinfo {year} {2021})},\ \Eprint {https://arxiv.org/abs/2103.06970} {arXiv:2103.06970 [quant-ph]} \BibitemShut {NoStop}%
\bibitem [{\citenamefont {Jayasinghe}\ \emph {et~al.}(2017)\citenamefont {Jayasinghe}, \citenamefont {Markantonakis}, \citenamefont {Akram},\ and\ \citenamefont {Mayes}}]{Danushka.2017}%
  \BibitemOpen
  \bibfield  {author} {\bibinfo {author} {\bibfnamefont {D.}~\bibnamefont {Jayasinghe}}, \bibinfo {author} {\bibfnamefont {K.}~\bibnamefont {Markantonakis}}, \bibinfo {author} {\bibfnamefont {R.~N.}\ \bibnamefont {Akram}},\ and\ \bibinfo {author} {\bibfnamefont {K.}~\bibnamefont {Mayes}},\ }\bibfield  {title} {\bibinfo {title} {Enhancing emv tokenisation with dynamic transaction tokens},\ }\href {https://link.springer.com/chapter/10.1007/978-3-319-62024-4_8} {\bibfield  {journal} {\bibinfo  {journal} {Radio Frequency Identification and IoT Security}\ } (\bibinfo {year} {2017})}\BibitemShut {NoStop}%
\bibitem [{\citenamefont {Amiri}\ \emph {et~al.}(2016)\citenamefont {Amiri}, \citenamefont {Wallden}, \citenamefont {Kent},\ and\ \citenamefont {Andersson}}]{AWK:PRA16}%
  \BibitemOpen
  \bibfield  {author} {\bibinfo {author} {\bibfnamefont {R.}~\bibnamefont {Amiri}}, \bibinfo {author} {\bibfnamefont {P.}~\bibnamefont {Wallden}}, \bibinfo {author} {\bibfnamefont {A.}~\bibnamefont {Kent}},\ and\ \bibinfo {author} {\bibfnamefont {E.}~\bibnamefont {Andersson}},\ }\bibfield  {title} {\bibinfo {title} {Secure quantum signatures using insecure quantum channels},\ }\href {https://doi.org/10.1103/PhysRevA.93.032325} {\bibfield  {journal} {\bibinfo  {journal} {Phys. Rev. A}\ }\textbf {\bibinfo {volume} {93}},\ \bibinfo {pages} {032325} (\bibinfo {year} {2016})},\ \Eprint {https://arxiv.org/abs/1507.02975} {arXiv:1507.02975 [quant-ph]} \BibitemShut {NoStop}%
\bibitem [{\citenamefont {Yin}\ \emph {et~al.}(2022)\citenamefont {Yin}, \citenamefont {Fu}, \citenamefont {Li}, \citenamefont {Weng}, \citenamefont {Li}, \citenamefont {Gu}, \citenamefont {Lu}, \citenamefont {Huang},\ and\ \citenamefont {Chen}}]{YYC:NSR22}%
  \BibitemOpen
  \bibfield  {author} {\bibinfo {author} {\bibfnamefont {H.-L.}\ \bibnamefont {Yin}}, \bibinfo {author} {\bibfnamefont {Y.}~\bibnamefont {Fu}}, \bibinfo {author} {\bibfnamefont {C.-L.}\ \bibnamefont {Li}}, \bibinfo {author} {\bibfnamefont {C.-X.}\ \bibnamefont {Weng}}, \bibinfo {author} {\bibfnamefont {B.-H.}\ \bibnamefont {Li}}, \bibinfo {author} {\bibfnamefont {J.}~\bibnamefont {Gu}}, \bibinfo {author} {\bibfnamefont {Y.-S.}\ \bibnamefont {Lu}}, \bibinfo {author} {\bibfnamefont {S.}~\bibnamefont {Huang}},\ and\ \bibinfo {author} {\bibfnamefont {Z.-B.}\ \bibnamefont {Chen}},\ }\bibfield  {title} {\bibinfo {title} {{Experimental quantum secure network with digital signatures and encryption}},\ }\href {https://doi.org/10.1093/nsr/nwac228} {\bibfield  {journal} {\bibinfo  {journal} {Natl. Sci. Rev.}\ } (\bibinfo {year} {2022})},\ \Eprint {https://arxiv.org/abs/2107.14089} {arXiv:2107.14089 [quant-ph]} \BibitemShut {NoStop}%
\bibitem [{\citenamefont {{Gilbert, Edgar N and MacWilliams, F Jessie and Sloane, Neil JA}}(1974)}]{gilbert1974codes}%
  \BibitemOpen
  \bibfield  {author} {\bibinfo {author} {\bibnamefont {{Gilbert, Edgar N and MacWilliams, F Jessie and Sloane, Neil JA}}},\ }\bibfield  {title} {\bibinfo {title} {{Codes which detect deception}},\ }\href {https://ieeexplore.ieee.org/document/6770988} {\bibfield  {journal} {\bibinfo  {journal} {Bell System Technical Journal}\ }\textbf {\bibinfo {volume} {53}},\ \bibinfo {pages} {405} (\bibinfo {year} {1974})}\BibitemShut {NoStop}%
\bibitem [{\citenamefont {Fak}(1979)}]{Fak.1979}%
  \BibitemOpen
  \bibfield  {author} {\bibinfo {author} {\bibfnamefont {V.}~\bibnamefont {Fak}},\ }\bibfield  {title} {\bibinfo {title} {{Repeated use of codes which detect deception (Corresp.)}},\ }\href {https://doi.org/10.1109/TIT.1979.1056011} {\bibfield  {journal} {\bibinfo  {journal} {{IEEE Transactions on Information Theory}}\ }\textbf {\bibinfo {volume} {25}},\ \bibinfo {pages} {233} (\bibinfo {year} {1979})}\BibitemShut {NoStop}%
\bibitem [{\citenamefont {{Rosenbaum, Ute}}(1993)}]{rosenbaum1993lower}%
  \BibitemOpen
  \bibfield  {author} {\bibinfo {author} {\bibnamefont {{Rosenbaum, Ute}}},\ }\bibfield  {title} {\bibinfo {title} {{A lower bound on authentication after having observed a sequence of messages}},\ }\href {https://link.springer.com/article/10.1007/BF00198462} {\bibfield  {journal} {\bibinfo  {journal} {{Journal of Cryptology}}\ }\textbf {\bibinfo {volume} {6}},\ \bibinfo {pages} {135} (\bibinfo {year} {1993})}\BibitemShut {NoStop}%
\bibitem [{\citenamefont {Wegman}\ and\ \citenamefont {Carter}(1981)}]{Wegman.1981}%
  \BibitemOpen
  \bibfield  {author} {\bibinfo {author} {\bibfnamefont {M.~N.}\ \bibnamefont {Wegman}}\ and\ \bibinfo {author} {\bibfnamefont {J.}~\bibnamefont {Carter}},\ }\bibfield  {title} {\bibinfo {title} {{New hash functions and their use in authentication and set equality}},\ }\href {https://doi.org/10.1016/0022-0000(81)90033-7} {\bibfield  {journal} {\bibinfo  {journal} {{Journal of Computer and System Sciences}}\ }\textbf {\bibinfo {volume} {22}},\ \bibinfo {pages} {265} (\bibinfo {year} {1981})}\BibitemShut {NoStop}%
\bibitem [{\citenamefont {Ghosh}\ and\ \citenamefont {Sarkar}(2021)}]{Ghosh.2021}%
  \BibitemOpen
  \bibfield  {author} {\bibinfo {author} {\bibfnamefont {S.}~\bibnamefont {Ghosh}}\ and\ \bibinfo {author} {\bibfnamefont {P.}~\bibnamefont {Sarkar}},\ }\bibfield  {title} {\bibinfo {title} {{Variants of Wegman-Carter message authentication code supporting variable tag lengths}},\ }\href {https://doi.org/10.1007/s10623-020-00840-w} {\bibfield  {journal} {\bibinfo  {journal} {{Designs, Codes and Cryptography}}\ }\textbf {\bibinfo {volume} {89}},\ \bibinfo {pages} {709} (\bibinfo {year} {2021})}\BibitemShut {NoStop}%
\bibitem [{\citenamefont {Diffie}\ and\ \citenamefont {Hellman}(1976)}]{Diffie}%
  \BibitemOpen
  \bibfield  {author} {\bibinfo {author} {\bibfnamefont {W.}~\bibnamefont {Diffie}}\ and\ \bibinfo {author} {\bibfnamefont {M.}~\bibnamefont {Hellman}},\ }\bibfield  {title} {\bibinfo {title} {New directions in cryptography},\ }\href {https://doi.org/10.1109/TIT.1976.1055638} {\bibfield  {journal} {\bibinfo  {journal} {IEEE Transactions on Information Theory}\ }\textbf {\bibinfo {volume} {22}},\ \bibinfo {pages} {644} (\bibinfo {year} {1976})}\BibitemShut {NoStop}%
\bibitem [{\citenamefont {Lo}\ and\ \citenamefont {Chau}(1997)}]{LC:PRL97}%
  \BibitemOpen
  \bibfield  {author} {\bibinfo {author} {\bibfnamefont {H.-K.}\ \bibnamefont {Lo}}\ and\ \bibinfo {author} {\bibfnamefont {H.~F.}\ \bibnamefont {Chau}},\ }\bibfield  {title} {\bibinfo {title} {Is quantum bit commitment really possible?},\ }\href {https://doi.org/10.1103/PhysRevLett.78.3410} {\bibfield  {journal} {\bibinfo  {journal} {Phys. Rev. Lett.}\ }\textbf {\bibinfo {volume} {78}},\ \bibinfo {pages} {3410} (\bibinfo {year} {1997})},\ \Eprint {https://arxiv.org/abs/9603004} {arXiv:9603004 [quant-ph]} \BibitemShut {NoStop}%
\bibitem [{\citenamefont {Mayers}(1997)}]{M:PRL97}%
  \BibitemOpen
  \bibfield  {author} {\bibinfo {author} {\bibfnamefont {D.}~\bibnamefont {Mayers}},\ }\bibfield  {title} {\bibinfo {title} {Unconditionally secure quantum bit commitment is impossible},\ }\href {https://doi.org/10.1103/PhysRevLett.78.3414} {\bibfield  {journal} {\bibinfo  {journal} {Phys. Rev. Lett.}\ }\textbf {\bibinfo {volume} {78}},\ \bibinfo {pages} {3414} (\bibinfo {year} {1997})},\ \Eprint {https://arxiv.org/abs/quant-ph/9605044} {arXiv:quant-ph/9605044 [quant-ph]} \BibitemShut {NoStop}%
\bibitem [{\citenamefont {{Neumann}}\ \emph {et~al.}(2022)\citenamefont {{Neumann}}, \citenamefont {{Selimovic}}, \citenamefont {{Bohmann}},\ and\ \citenamefont {{Ursin}}}]{Ursin2021gbitsource}%
  \BibitemOpen
  \bibfield  {author} {\bibinfo {author} {\bibfnamefont {S.~P.}\ \bibnamefont {{Neumann}}}, \bibinfo {author} {\bibfnamefont {M.}~\bibnamefont {{Selimovic}}}, \bibinfo {author} {\bibfnamefont {M.}~\bibnamefont {{Bohmann}}},\ and\ \bibinfo {author} {\bibfnamefont {R.}~\bibnamefont {{Ursin}}},\ }\bibfield  {title} {\bibinfo {title} {{Experimental entanglement generation for quantum key distribution beyond 1 Gbit/s}},\ }\href {https://doi.org/10.22331/q-2022-09-29-822} {\bibfield  {journal} {\bibinfo  {journal} {Quantum}\ }\textbf {\bibinfo {volume} {6}},\ \bibinfo {pages} {822} (\bibinfo {year} {2022})},\ \Eprint {https://arxiv.org/abs/2107.07756} {arXiv:2107.07756 [quant-ph]} \BibitemShut {NoStop}%
\bibitem [{\citenamefont {Gyongyosi}\ and\ \citenamefont {Imre}(2022)}]{Gyo:cacm22}%
  \BibitemOpen
  \bibfield  {author} {\bibinfo {author} {\bibfnamefont {L.}~\bibnamefont {Gyongyosi}}\ and\ \bibinfo {author} {\bibfnamefont {S.}~\bibnamefont {Imre}},\ }\bibfield  {title} {\bibinfo {title} {Advances in the quantum internet},\ }\href {https://doi.org/10.1145/3524455} {\bibfield  {journal} {\bibinfo  {journal} {Commun. ACM}\ }\textbf {\bibinfo {volume} {65}},\ \bibinfo {pages} {52–63} (\bibinfo {year} {2022})}\BibitemShut {NoStop}%
\bibitem [{\citenamefont {Dubhashi}\ and\ \citenamefont {Panconesi}(2009)}]{DP09}%
  \BibitemOpen
  \bibfield  {author} {\bibinfo {author} {\bibfnamefont {D.~P.}\ \bibnamefont {Dubhashi}}\ and\ \bibinfo {author} {\bibfnamefont {A.}~\bibnamefont {Panconesi}},\ }\bibfield  {title} {\bibinfo {title} {Concentration of measure for the analysis of randomized algorithms},\ }\href {https://doi.org/10.1017/CBO9780511581274} {\bibfield  {journal} {\bibinfo  {journal} {Concentration of Measure for the Analysis of Randomized Algorithms}\ } (\bibinfo {year} {2009})}\BibitemShut {NoStop}%
\bibitem [{\citenamefont {Vandenberghe}\ and\ \citenamefont {Boyd}(1996)}]{VB:SIAM96}%
  \BibitemOpen
  \bibfield  {author} {\bibinfo {author} {\bibfnamefont {L.}~\bibnamefont {Vandenberghe}}\ and\ \bibinfo {author} {\bibfnamefont {S.}~\bibnamefont {Boyd}},\ }\bibfield  {title} {\bibinfo {title} {Semidefinite programming},\ }\href {https://doi.org/10.1137/1038003} {\bibfield  {journal} {\bibinfo  {journal} {SIAM Review}\ }\textbf {\bibinfo {volume} {38}},\ \bibinfo {pages} {49} (\bibinfo {year} {1996})}\BibitemShut {NoStop}%
\bibitem [{\citenamefont {Watrous}(2011)}]{W:LN11}%
  \BibitemOpen
  \bibfield  {author} {\bibinfo {author} {\bibfnamefont {J.}~\bibnamefont {Watrous}},\ }\bibfield  {title} {\bibinfo {title} {Semidefinite programming},\ }\href {https://web.archive.org/web/20220521201106/https://cs.uwaterloo.ca/~watrous/TQI-notes/} {\bibfield  {journal} {\bibinfo  {journal} {Theory of Quantum Information (notes from Fall 2011)}\ } (\bibinfo {year} {2011})}\BibitemShut {NoStop}%
\bibitem [{\citenamefont {Bra{\'{n}}czyk}\ \emph {et~al.}(2010)\citenamefont {Bra{\'{n}}czyk}, \citenamefont {Ralph}, \citenamefont {Helwig},\ and\ \citenamefont {Silberhorn}}]{BR:NJP10}%
  \BibitemOpen
  \bibfield  {author} {\bibinfo {author} {\bibfnamefont {A.~M.}\ \bibnamefont {Bra{\'{n}}czyk}}, \bibinfo {author} {\bibfnamefont {T.~C.}\ \bibnamefont {Ralph}}, \bibinfo {author} {\bibfnamefont {W.}~\bibnamefont {Helwig}},\ and\ \bibinfo {author} {\bibfnamefont {C.}~\bibnamefont {Silberhorn}},\ }\bibfield  {title} {\bibinfo {title} {Optimized generation of heralded fock states using parametric down-conversion},\ }\href {https://doi.org/10.1088/1367-2630/12/6/063001} {\bibfield  {journal} {\bibinfo  {journal} {New J. Phys.}\ }\textbf {\bibinfo {volume} {12}},\ \bibinfo {pages} {063001} (\bibinfo {year} {2010})},\ \Eprint {https://arxiv.org/abs/0909.4147} {arXiv:0909.4147 [quant-ph]} \BibitemShut {NoStop}%
\bibitem [{\citenamefont {Laudenbach}\ \emph {et~al.}(2017)\citenamefont {Laudenbach}, \citenamefont {Kalista}, \citenamefont {Hentschel}, \citenamefont {Walther},\ and\ \citenamefont {Hübel}}]{LKH:SciRep17}%
  \BibitemOpen
  \bibfield  {author} {\bibinfo {author} {\bibfnamefont {F.}~\bibnamefont {Laudenbach}}, \bibinfo {author} {\bibfnamefont {S.}~\bibnamefont {Kalista}}, \bibinfo {author} {\bibfnamefont {M.}~\bibnamefont {Hentschel}}, \bibinfo {author} {\bibfnamefont {P.}~\bibnamefont {Walther}},\ and\ \bibinfo {author} {\bibfnamefont {H.}~\bibnamefont {Hübel}},\ }\bibfield  {title} {\bibinfo {title} {A novel single-crystal \& single-pass source for polarisation- and colour-entangled photon pairs},\ }\href {https://doi.org/10.1038/s41598-017-07781-w} {\bibfield  {journal} {\bibinfo  {journal} {Sci. Rep.}\ }\textbf {\bibinfo {volume} {7}},\ \bibinfo {pages} {7235} (\bibinfo {year} {2017})},\ \Eprint {https://arxiv.org/abs/1612.06579} {arXiv:1612.06579 [quant-ph]} \BibitemShut {NoStop}%
\bibitem [{\citenamefont {Roehsner}\ \emph {et~al.}(2021)\citenamefont {Roehsner}, \citenamefont {Kettlewell}, \citenamefont {Fitzsimons},\ and\ \citenamefont {Walther}}]{RKF:npj21}%
  \BibitemOpen
  \bibfield  {author} {\bibinfo {author} {\bibfnamefont {M.-C.}\ \bibnamefont {Roehsner}}, \bibinfo {author} {\bibfnamefont {J.~A.}\ \bibnamefont {Kettlewell}}, \bibinfo {author} {\bibfnamefont {J.}~\bibnamefont {Fitzsimons}},\ and\ \bibinfo {author} {\bibfnamefont {P.}~\bibnamefont {Walther}},\ }\bibfield  {title} {\bibinfo {title} {Probabilistic one-time programs using quantum entanglement},\ }\href {https://doi.org/10.1038/s41534-021-00435-w} {\bibfield  {journal} {\bibinfo  {journal} {npj Quantum Inf.}\ }\textbf {\bibinfo {volume} {7}},\ \bibinfo {pages} {98} (\bibinfo {year} {2021})},\ \Eprint {https://arxiv.org/abs/2008.02294} {arXiv:2008.02294 [quant-ph]} \BibitemShut {NoStop}%
\bibitem [{\citenamefont {{Signorini}}\ \emph {et~al.}(2021)\citenamefont {{Signorini}}, \citenamefont {{Sanna}}, \citenamefont {{Piccione}}, \citenamefont {{Ghulinyan}}, \citenamefont {{Tidemand-Lichtenberg}}, \citenamefont {{Pedersen}},\ and\ \citenamefont {{Pavesi}}}]{signorini2021}%
  \BibitemOpen
  \bibfield  {author} {\bibinfo {author} {\bibfnamefont {S.}~\bibnamefont {{Signorini}}}, \bibinfo {author} {\bibfnamefont {M.}~\bibnamefont {{Sanna}}}, \bibinfo {author} {\bibfnamefont {S.}~\bibnamefont {{Piccione}}}, \bibinfo {author} {\bibfnamefont {M.}~\bibnamefont {{Ghulinyan}}}, \bibinfo {author} {\bibfnamefont {P.}~\bibnamefont {{Tidemand-Lichtenberg}}}, \bibinfo {author} {\bibfnamefont {C.}~\bibnamefont {{Pedersen}}},\ and\ \bibinfo {author} {\bibfnamefont {L.}~\bibnamefont {{Pavesi}}},\ }\bibfield  {title} {\bibinfo {title} {{A silicon source of heralded single photons at 2 {\ensuremath{\mu}}m}},\ }\href {https://doi.org/10.1063/5.0063393} {\bibfield  {journal} {\bibinfo  {journal} {APL Photonics}\ }\textbf {\bibinfo {volume} {6}},\ \bibinfo {pages} {126103} (\bibinfo {year} {2021})},\ \Eprint {https://arxiv.org/abs/2108.01031} {arXiv:2108.01031 [quant-ph]} \BibitemShut {NoStop}%
\end{thebibliography}%


\begin{thebibliography}{14}%
\makeatletter
\providecommand \@ifxundefined [1]{%
 \@ifx{#1\undefined}
}%
\providecommand \@ifnum [1]{%
 \ifnum #1\expandafter \@firstoftwo
 \else \expandafter \@secondoftwo
 \fi
}%
\providecommand \@ifx [1]{%
 \ifx #1\expandafter \@firstoftwo
 \else \expandafter \@secondoftwo
 \fi
}%
\providecommand \natexlab [1]{#1}%
\providecommand \enquote  [1]{``#1''}%
\providecommand \bibnamefont  [1]{#1}%
\providecommand \bibfnamefont [1]{#1}%
\providecommand \citenamefont [1]{#1}%
\providecommand \href@noop [0]{\@secondoftwo}%
\providecommand \href [0]{\begingroup \@sanitize@url \@href}%
\providecommand \@href[1]{\@@startlink{#1}\@@href}%
\providecommand \@@href[1]{\endgroup#1\@@endlink}%
\providecommand \@sanitize@url [0]{\catcode `\\12\catcode `\$12\catcode `\&12\catcode `\#12\catcode `\^12\catcode `\_12\catcode `\%12\relax}%
\providecommand \@@startlink[1]{}%
\providecommand \@@endlink[0]{}%
\providecommand \url  [0]{\begingroup\@sanitize@url \@url }%
\providecommand \@url [1]{\endgroup\@href {#1}{\urlprefix }}%
\providecommand \urlprefix  [0]{URL }%
\providecommand \Eprint [0]{\href }%
\providecommand \doibase [0]{https://doi.org/}%
\providecommand \selectlanguage [0]{\@gobble}%
\providecommand \bibinfo  [0]{\@secondoftwo}%
\providecommand \bibfield  [0]{\@secondoftwo}%
\providecommand \translation [1]{[#1]}%
\providecommand \BibitemOpen [0]{}%
\providecommand \bibitemStop [0]{}%
\providecommand \bibitemNoStop [0]{.\EOS\space}%
\providecommand \EOS [0]{\spacefactor3000\relax}%
\providecommand \BibitemShut  [1]{\csname bibitem#1\endcsname}%
\let\auto@bib@innerbib\@empty
\bibitem [{\citenamefont {Watrous}(2011)}]{W:LN11}%
  \BibitemOpen
  \bibfield  {author} {\bibinfo {author} {\bibfnamefont {J.}~\bibnamefont {Watrous}},\ }\bibfield  {title} {\bibinfo {title} {Semidefinite programming},\ }\href {https://web.archive.org/web/20220521201106/https://cs.uwaterloo.ca/~watrous/TQI-notes/} {\bibfield  {journal} {\bibinfo  {journal} {Theory of Quantum Information (notes from Fall 2011)}\ } (\bibinfo {year} {2011})}\BibitemShut {NoStop}%
\bibitem [{\citenamefont {Vandenberghe}\ and\ \citenamefont {Boyd}(1996)}]{VB:SIAM96}%
  \BibitemOpen
  \bibfield  {author} {\bibinfo {author} {\bibfnamefont {L.}~\bibnamefont {Vandenberghe}}\ and\ \bibinfo {author} {\bibfnamefont {S.}~\bibnamefont {Boyd}},\ }\bibfield  {title} {\bibinfo {title} {Semidefinite programming},\ }\href {https://doi.org/10.1137/1038003} {\bibfield  {journal} {\bibinfo  {journal} {SIAM Review}\ }\textbf {\bibinfo {volume} {38}},\ \bibinfo {pages} {49} (\bibinfo {year} {1996})}\BibitemShut {NoStop}%
\bibitem [{\citenamefont {Molina}\ \emph {et~al.}(2013)\citenamefont {Molina}, \citenamefont {Vidick},\ and\ \citenamefont {Watrous}}]{MVW:tqc12}%
  \BibitemOpen
  \bibfield  {author} {\bibinfo {author} {\bibfnamefont {A.}~\bibnamefont {Molina}}, \bibinfo {author} {\bibfnamefont {T.}~\bibnamefont {Vidick}},\ and\ \bibinfo {author} {\bibfnamefont {J.}~\bibnamefont {Watrous}},\ }\bibfield  {title} {\bibinfo {title} {Optimal counterfeiting attacks and generalizations for wiesner's quantum money}\ }(\bibinfo  {publisher} {Springer},\ \bibinfo {year} {2013})\BibitemShut {NoStop}%
\bibitem [{\citenamefont {Chernoff}(1952)}]{Chernoff:52}%
  \BibitemOpen
  \bibfield  {author} {\bibinfo {author} {\bibfnamefont {H.}~\bibnamefont {Chernoff}},\ }\bibfield  {title} {\bibinfo {title} {{A Measure of Asymptotic Efficiency for Tests of a Hypothesis Based on the sum of Observations}},\ }\href {https://doi.org/10.1214/aoms/1177729330} {\bibfield  {journal} {\bibinfo  {journal} {The Annals of Mathematical Statistics}\ }\textbf {\bibinfo {volume} {23}},\ \bibinfo {pages} {493 } (\bibinfo {year} {1952})}\BibitemShut {NoStop}%
\bibitem [{\citenamefont {Hoeffding}(1994)}]{Hoeffding1994}%
  \BibitemOpen
  \bibfield  {author} {\bibinfo {author} {\bibfnamefont {W.}~\bibnamefont {Hoeffding}},\ }\bibinfo {title} {Probability inequalities for sums of bounded random variables},\ in\ \href {https://doi.org/10.1007/978-1-4612-0865-5_26} {\emph {\bibinfo {booktitle} {The Collected Works of Wassily Hoeffding}}},\ \bibinfo {editor} {edited by\ \bibinfo {editor} {\bibfnamefont {N.~I.}\ \bibnamefont {Fisher}}\ and\ \bibinfo {editor} {\bibfnamefont {P.~K.}\ \bibnamefont {Sen}}}\ (\bibinfo  {publisher} {Springer New York},\ \bibinfo {address} {New York, NY},\ \bibinfo {year} {1994})\ pp.\ \bibinfo {pages} {409--426}\BibitemShut {NoStop}%
\bibitem [{\citenamefont {Mitzenmacher}\ and\ \citenamefont {Upfal}(2005)}]{M:2005}%
  \BibitemOpen
  \bibfield  {author} {\bibinfo {author} {\bibfnamefont {M.}~\bibnamefont {Mitzenmacher}}\ and\ \bibinfo {author} {\bibfnamefont {E.}~\bibnamefont {Upfal}},\ }\href {https://doi.org/10.1017/CBO9780511813603} {\emph {\bibinfo {title} {Probability and Computing: Randomized Algorithms and Probabilistic Analysis}}}\ (\bibinfo  {publisher} {Cambridge University Press},\ \bibinfo {year} {2005})\BibitemShut {NoStop}%
\bibitem [{\citenamefont {Katz}\ and\ \citenamefont {Lindell}(2014)}]{Katz.2014}%
  \BibitemOpen
  \bibfield  {author} {\bibinfo {author} {\bibfnamefont {J.}~\bibnamefont {Katz}}\ and\ \bibinfo {author} {\bibfnamefont {Y.}~\bibnamefont {Lindell}},\ }\href {https://doi.org/10.1201/b17668} {\emph {\bibinfo {title} {{Introduction to Modern Cryptography, Second Edition}}}}\ (\bibinfo  {publisher} {{Chapman and Hall/CRC}},\ \bibinfo {year} {2014})\BibitemShut {NoStop}%
\bibitem [{\citenamefont {Stinson}(2006)}]{Stinson.2006}%
  \BibitemOpen
  \bibfield  {author} {\bibinfo {author} {\bibfnamefont {D.~R.}\ \bibnamefont {Stinson}},\ }\href@noop {} {\emph {\bibinfo {title} {{Cryptography: Theory and practice}}}},\ \bibinfo {edition} {3rd}\ ed.,\ {Discrete mathematics and its applications}\ (\bibinfo  {publisher} {{Chapman {\&} Hall/CRC}},\ \bibinfo {address} {Londen [etc.]},\ \bibinfo {year} {2006})\BibitemShut {NoStop}%
\bibitem [{\citenamefont {Wegman}\ and\ \citenamefont {Carter}(1981)}]{Wegman.1981}%
  \BibitemOpen
  \bibfield  {author} {\bibinfo {author} {\bibfnamefont {M.~N.}\ \bibnamefont {Wegman}}\ and\ \bibinfo {author} {\bibfnamefont {J.}~\bibnamefont {Carter}},\ }\bibfield  {title} {\bibinfo {title} {{New hash functions and their use in authentication and set equality}},\ }\href {https://doi.org/10.1016/0022-0000(81)90033-7} {\bibfield  {journal} {\bibinfo  {journal} {{Journal of Computer and System Sciences}}\ }\textbf {\bibinfo {volume} {22}},\ \bibinfo {pages} {265} (\bibinfo {year} {1981})}\BibitemShut {NoStop}%
\bibitem [{\citenamefont {{Gilbert, Edgar N and MacWilliams, F Jessie and Sloane, Neil JA}}(1974)}]{gilbert1974codes}%
  \BibitemOpen
  \bibfield  {author} {\bibinfo {author} {\bibnamefont {{Gilbert, Edgar N and MacWilliams, F Jessie and Sloane, Neil JA}}},\ }\bibfield  {title} {\bibinfo {title} {{Codes which detect deception}},\ }\href {https://ieeexplore.ieee.org/document/6770988} {\bibfield  {journal} {\bibinfo  {journal} {Bell System Technical Journal}\ }\textbf {\bibinfo {volume} {53}},\ \bibinfo {pages} {405} (\bibinfo {year} {1974})}\BibitemShut {NoStop}%
\bibitem [{\citenamefont {Fak}(1979)}]{Fak.1979}%
  \BibitemOpen
  \bibfield  {author} {\bibinfo {author} {\bibfnamefont {V.}~\bibnamefont {Fak}},\ }\bibfield  {title} {\bibinfo {title} {{Repeated use of codes which detect deception (Corresp.)}},\ }\href {https://doi.org/10.1109/TIT.1979.1056011} {\bibfield  {journal} {\bibinfo  {journal} {{IEEE Transactions on Information Theory}}\ }\textbf {\bibinfo {volume} {25}},\ \bibinfo {pages} {233} (\bibinfo {year} {1979})}\BibitemShut {NoStop}%
\bibitem [{\citenamefont {{Rosenbaum, Ute}}(1993)}]{rosenbaum1993lower}%
  \BibitemOpen
  \bibfield  {author} {\bibinfo {author} {\bibnamefont {{Rosenbaum, Ute}}},\ }\bibfield  {title} {\bibinfo {title} {{A lower bound on authentication after having observed a sequence of messages}},\ }\href {https://link.springer.com/article/10.1007/BF00198462} {\bibfield  {journal} {\bibinfo  {journal} {{Journal of Cryptology}}\ }\textbf {\bibinfo {volume} {6}},\ \bibinfo {pages} {135} (\bibinfo {year} {1993})}\BibitemShut {NoStop}%
\bibitem [{\citenamefont {Bra{\'{n}}czyk}\ \emph {et~al.}(2010)\citenamefont {Bra{\'{n}}czyk}, \citenamefont {Ralph}, \citenamefont {Helwig},\ and\ \citenamefont {Silberhorn}}]{BR:NJP10}%
  \BibitemOpen
  \bibfield  {author} {\bibinfo {author} {\bibfnamefont {A.~M.}\ \bibnamefont {Bra{\'{n}}czyk}}, \bibinfo {author} {\bibfnamefont {T.~C.}\ \bibnamefont {Ralph}}, \bibinfo {author} {\bibfnamefont {W.}~\bibnamefont {Helwig}},\ and\ \bibinfo {author} {\bibfnamefont {C.}~\bibnamefont {Silberhorn}},\ }\bibfield  {title} {\bibinfo {title} {Optimized generation of heralded fock states using parametric down-conversion},\ }\href {https://doi.org/10.1088/1367-2630/12/6/063001} {\bibfield  {journal} {\bibinfo  {journal} {New J. Phys.}\ }\textbf {\bibinfo {volume} {12}},\ \bibinfo {pages} {063001} (\bibinfo {year} {2010})},\ \Eprint {https://arxiv.org/abs/0909.4147} {arXiv:0909.4147 [quant-ph]} \BibitemShut {NoStop}%
\bibitem [{\citenamefont {Renner}(2005)}]{Renner:Finetti}%
  \BibitemOpen
  \bibfield  {author} {\bibinfo {author} {\bibfnamefont {R.}~\bibnamefont {Renner}},\ }\href {https://arxiv.org/abs/quant-ph/0512258} {\bibinfo {title} {Security of quantum key distribution}} (\bibinfo {year} {2005}),\ \Eprint {https://arxiv.org/abs/0512258} {arXiv:0512258 [quant-ph]} \BibitemShut {NoStop}%
\end{thebibliography}%

\end{document}


\title{Demonstration of quantum-digital payments - Supplementary material}

\author{Peter Schiansky}
\thanks{These two authors contributed equally.}
\affiliation{University of Vienna, Faculty of Physics, Vienna Center for Quantum Science and Technology (VCQ), 1090 Vienna, Austria}

\author{Julia Kalb}
\thanks{These two authors contributed equally.}
\affiliation{University of Vienna, Faculty of Physics, Vienna Center for Quantum Science and Technology (VCQ), 1090 Vienna, Austria}

\author{Esther Sztatecsny}
\affiliation{University of Vienna, Faculty of Physics, Vienna Center for Quantum Science and Technology (VCQ), 1090 Vienna, Austria}

\author{Marie-Christine Roehsner}
\thanks{Current address: QuTech \& Kavli Institute of Nanoscience, Delft University of Technology, 2628 CJ Delft, The Netherlands.}
\affiliation{University of Vienna, Faculty of Physics, Vienna Center for Quantum Science and Technology (VCQ), 1090 Vienna, Austria}

\author{Tobias Guggemos}
\affiliation{University of Vienna, Faculty of Physics, Vienna Center for Quantum Science and Technology (VCQ), 1090 Vienna, Austria}

\author{Alessandro Trenti}
\thanks{Current address: Security and Communication Technologies, Center for Digital Safety and Security, AIT Austrian Institute of Technology GmbH, Giefinggasse 4, 1210 Vienna, Austria.}
\affiliation{University of Vienna, Faculty of Physics, Vienna Center for Quantum Science and Technology (VCQ), 1090 Vienna, Austria}

\author{Mathieu Bozzio}\email[Corresponding author: ]{mathieu.bozzio@univie.ac.at}
\affiliation{University of Vienna, Faculty of Physics, Vienna Center for Quantum Science and Technology (VCQ), 1090 Vienna, Austria}

\author{Philip Walther}\email[Corresponding author: ]{philip.walther@univie.ac.at}
\affiliation{University of Vienna, Faculty of Physics, Vienna Center for Quantum Science and Technology (VCQ), 1090 Vienna, Austria}
\affiliation{Christian Doppler Laboratory for Photonic Quantum Computer, Faculty of Physics, University of Vienna, 1090 Vienna, Austria}

\usetikzlibrary{arrows.meta}
\tikzsymbolsset{after-symbol={}}
\tikzsymbolsdefinesymbol {hvbas}{ S } {%
    \begin{tikzpicture}[/tikzsymbolsstyle, x=1.1ex,y=1.1ex, line width=0.07ex*\tikzsymbolsscaleabs]
    \draw[{Latex[scale=0.5]}-{Latex[scale=0.5]}] (0,-1) -- (0,1);
    \draw[{Latex[scale=0.5]}-{Latex[scale=0.5]}] (-1,0) -- (1,0);
    \end{tikzpicture}
}
\tikzsymbolsdefinesymbol {adbas}{ S }{%
    \begin{tikzpicture}[/tikzsymbolsstyle, x=1.1ex,y=1.1ex, scale=#1, line width=0.07ex*\tikzsymbolsscaleabs]
    \draw[{Latex[scale=0.5]}-{Latex[scale=0.5]}] (-1*0.707,-1*0.707) -- (1*0.707,1*0.707);
    \draw[{Latex[scale=0.5]}-{Latex[scale=0.5]}] (-1*0.707,1*0.707) -- (1*0.707,-1*0.707);
    \end{tikzpicture}
}
\newacro{gps}[GPS]{Global Positioning System}
\newacro{hmac}[MAC]{Message Authentication Code}
\newacro{itsec}[i.t.-security]{information-theoretic security}
\newacroplural{itsec}[i.t.-secure]{information-theoretic secure}
\newacro{qkd}[QKD]{Quantum Key Distribution}
\newacro{spdc}[SPDC]{Spontaneous Parametric Down Conversion}
\newacro{ttm}[TTM]{Time-Tagging Modules}
\newacro{ttp}[TTP]{Trusted Token Provider}

\maketitle

\noindent We first introduce the mathematical tools required to understand our security analysis, namely semidefinite programming, Choi's theorem on completely positive maps and Chernoff-Hoeffding bounds. We then introduce the cryptographic tools required to understand the classical security analysis, namely hash functions and message authentication codes (MAC and HMAC), as well as their security properties. We then derive the information-theoretic security against double-spending, and prove the information-theoretic security of our cryptogram, concealing the secure key $C$. We finally provide our protocol performance along with detailed setup characterizations. 

\section{Mathematical preliminaries}\label{sec:prel}

\subsection{Semidefinite programming.}\label{sec:sdp}

\noindent Quantum theory relies on linear algebra. In quantum cryptography, security analyses often involve optimizing over semidefinite positive objects to find the adversary's optimal cheating strategy. Most of the time, these objects are density matrices, measurement operators, or more general completely positive trace-preserving (CPTP) maps. Semidefinite programming provides a suitable framework for this, as it allows to optimize over semidefinite positive variables, given linear constraints \cite{W:LN11,VB:SIAM96,MVW:tqc12}.

A semidefinite program may be defined as a triple $\left(\Lambda,F,C\right)$ where $\Lambda$ is a Hermitian-preserving CPTP map, and $F$ and $C$ are Hermitian operators living in complex Hilbert spaces $\mathcal{H}_F$ and $\mathcal{H}_C$, respectively. We start by defining a maximization problem, which will serve as our \textit{primal problem}. The primal problem maximizes a \textit{primal objective function}, $\Tr\left(F^{\dagger}X\right)$,  over all positive semidefinite variables $X$, given a set of linear constraints expressed as a function of $C$:

\begin{equation}
\begin{aligned}
\text{maximize} &&& \Tr\left(F^{\dagger}X\right)\\
\text{s.t.}  &&& \Lambda(X) = C\\
&&& X \geqslant 0.
\end{aligned}\label{primal}
\end{equation}
If it exists, the operator $X$ which maximizes $\Tr\left(F^{\dagger}X\right)$ given these constraints is the \textit{primal optimal solution}, and the corresponding value of $\Tr\left(F^{\dagger}X\right)$ is the  \textit{primal optimal value}.

Semidefinite programs present an elegant dual structure, which associates a dual minimization problem to each primal maximization problem. Effectively, the new variable(s) $Y$ of the dual problem may be understood as the Lagrange multipliers associated with the constraints of the primal problem (one for each constraint). The dual problem associated with \eqqref{primal} reads \cite{W:LN11,VB:SIAM96,MVW:tqc12}:

\begin{equation}
\begin{aligned}
\text{minimize} &&& \Tr\left(C^{\dagger}Y\right)\\
\text{s.t.}  &&& \Lambda^*(Y) - F \geqslant 0\\
&&& Y = Y^{\dagger}.
\end{aligned}\label{duall}
\end{equation}
 Similarly to the primal problem, the operator $Y$ which minimizes $\Tr\left(C^{\dagger}Y\right)$ given these constraints, if it exists, is the \textit{dual optimal solution}, and the corresponding value of $\Tr\left(C^{\dagger}Y\right)$ is the  \textit{dual optimal value}.

The Lagrange multiplier method allows to find the local extremum of a constrained function. The optimal value $s_p$ of the primal problem therefore lower bounds the optimal value $s_d$ of the dual problem, while the optimal value of the dual upper bounds that of the primal. This property is known as \textit{weak duality}, and may be simply expressed as:

\begin{equation}
    s_p\leqslant s_d.
\end{equation}
In many quantum-cryptographic applications however, we wish to ensure that the upper bound derived in the primal problem is \textit{tight}, i.e. that the local maximum is in fact a global maximum for the objective function. The dual problem will help to prove this when there exists \textit{strong duality}:

\begin{equation}
    s_p= s_d.
\end{equation}

\subsection{Choi's theorem on completely positive maps.}\label{sec:choi}

\noindent Let us consider a tensor product of two $d$-dimensional Hilbert spaces $\mathcal{H}=\mathcal{H}_1^d\otimes\mathcal{H}_2^d$, and then define the maximally entangled state $\ket{\Phi^{+}}\bra{\Phi^{+}}$ on $\mathcal{H}$ as:
\begin{equation}
\ket{\Phi^{+}}\bra{\Phi^{+}} = \frac{1}{d}\sum_{i,j=1}^{d} \ket{i}\bra{j}\otimes\ket{i}\bra{j}.
\end{equation}
We introduce a completely positive linear map $\Lambda : \mathcal{H}_1^d \rightarrow \mathcal{H}_3^{d'}$, and define the Choi-Jamiolkowski operator $J(\Lambda) : \mathcal{H}_1^d\otimes\mathcal{H}_2^d \rightarrow \mathcal{H}_3^{d'}\otimes\mathcal{H}_2^d$ as the operator which applies $\Lambda$ to the first half of the maximally entangled state $\ket{\Phi^{+}}\bra{\Phi^{+}}$:
\begin{equation}
J(\Lambda) = \frac{1}{d}\sum_{i,j=1}^{d} \Lambda(\ket{i}\bra{j})\otimes\ket{i}\bra{j}.
\label{eq:choi}
\end{equation}
Choi's theorem then states that $\Lambda$ is completely positive if and only if
$J(\Lambda)$ is positive semidefinite. We also have that $\Lambda$ is a trace-preserving map if and only if $\Tr_{\mathcal{H}_3^{d'}}(J(\Lambda)) = \mathbb{1}_{\mathcal{H}_2^d}$ \cite{W:LN11,VB:SIAM96,MVW:tqc12}.

\subsection{Chernoff-Hoeffding bounds.}\label{Chernoff_Hoeffding}

Chernoff-Hoeffding inequalities provide exponentially decreasing bounds on tail distributions of sums of independent random variables \cite{Chernoff:52,Hoeffding1994,M:2005}.

Let us assume a set of $N$ random variables $X_i$, which can each take the value $1$ with probability $p_i$ and $0$ with probability $1-p_i$. The probability that $X= \sum_{i=0}^{N-1}X_i$ is larger than the expected mean value $\mu$ by some $\delta \in [0,1]$ is bounded by the following \textit{upper tail}:
\begin{equation}\label{uppertail}
    P[X\geq (1+\delta)\cdot \mu] \leq \exp{\left(-\frac{\delta^2}{3}\cdot \mu\cdot N\right)}
\end{equation}
\noindent
Similarly, the probability that $X= \sum_{i=0}^{N-1}X_i$ is smaller than the expected mean value by some $\delta \in [0,1]$ is bounded by the \textit{lower tail}:

\begin{equation}\label{lowertail}
    P[X\leq (1-\delta)\cdot \mu] \leq \exp{\left(-\frac{\delta^2}{2}\cdot \mu\cdot N\right)}
\end{equation}

\section{Cryptographic preliminaries}\label{sec:crypto}

We first describe the preliminaries of hash functions, \textit{MAC}s and H\textit{MAC}s in a computationally-secure setting (such that attacks are bounded by polynomial time attackers). 
We then show how to construct an \textit{i.t.-secure} H\textit{MAC} with these ingredients and define its security bounds against \textit{unbounded} attackers. The interested reader is referred to~\cite{Katz.2014,Stinson.2006} for further reading.

\subsection{Hash function.}

A hash function is defined as a function that maps a set of arbitrary length to a finite set $H:\{0,1\}^*\mapsto\{0,1\}^n; n\in\mathbb{N}$.
Hence, a hash function  $ H $ is non-injective by definition, and threatened by:

\begin{itemize}[left=.5cm]
	\item \textbf{collision attacks:} identifying two different arbitrary inputs $ x_1 $ and $ x_2 $ such that $ H(x_1) = H(x_2) $,
	\item \textbf{second pre-image attacks:} finding a pre-image input $ x_2 $, such that $ H(x_1) = H(x_2) $,  given an input string $ x_1 $, 
	\item  \textbf{pre-image attacks:} for a given output string $ y $, finding a pre-image input $ x $, such that $ y = H(x) $.
\end{itemize}
If $ H $ withstands collision attacks, we call it \textit{collision-resistant}; in that case, $ H $ is also \textit{second pre-image} and \textit{pre-image} resistant.
We consider $ H $ to be a \textit{secure} or \textit{cryptographic} hash function, if it is \textit{collision-resistant} against computationally bounded attacks.

\subsection{Message Authentication Code (MAC).}
A \textit{MAC} is a function that takes a key $ k \in \{0,1\}^n; n \in \mathbb{N} $ and message $ m \in \{0,1\}^* $ as input and subsequently outputs a tag $ y \in \{0,1\}^* $ such that $ \mac(k,m) \mapsto y $.
Upon receiving the tuple $ (y,m) $, an honest verifier in possession of $ k $ can verify the authenticity of a message, i.e., $ y = \mac(k,m) $.

\smallskip

We consider a \textit{MAC} to be secure if \textit{existential unforgeability against a chosen message attack} holds, where:
\begin{itemize}[left=.5cm]
	\item\textbf{existentially unforgeable} means that the attacker is not able to generate a valid MAC tag on any message, without being in possession of the key $ k $,
	\item\textbf{chosen message attack} means that the attacker is able to obtain \textit{MAC} tags on any other messages before performing the attack.
\end{itemize}

\subsection{HMAC.}
A typical implementation of a \textit{MAC} is the so called \textit{Keyed-Hash Message Authentication Code (HMAC)}, which is a function $f(H,k,m) \mapsto y$.
Based on a hash function $H$, it takes a secret key $k$ and message $m$ as inputs, and outputs some authentication tag.
The $ \hmac $ under the use of the hash function $ H $ is defined as:
\[ \hmac(k,m) = H\big(k \oplus \text{opad} \parallel H\left(k\oplus \text{ipad} \parallel m\right)\big) ,\]
where $ \oplus $ is a bitwise XOR; $ \parallel $ means simple concatenation; and \texttt{opad} and \texttt{ipad} are fixed public strings with a length depending on the underlying function of $ H $.

We consider an \textit{HMAC} to be a secure MAC function (i.e., \textit{existentially unforgeable against a chosen message attack}), if the underlying hash function $ H $ is \textit{collision resistant}.

\subsection{i.t.-secure MAC.} \label{sec:hmac-it-secure}

A \textit{Message Authentication Code} (MAC) is called \textit{information-theoretically} or \textit{perfectly} or \textit{unconditionally} secure if it is secure against~\cite{Stinson.2006, Wegman.1981, gilbert1974codes}:
\begin{itemize}[left=.5cm]
	\item\textbf{impersonation attacks}: where an attacker can create a message and a tag, valid under the key in use,
	\item\textbf{substitution attacks}: where an attacker sees one valid message-tag pair, intercepts it, and then replaces it with another valid message-tag pair.
\end{itemize}
In other words, there is no $\mac(k,m) = \mac(k,m') $ for $ m \neq m' $ (given that $k$ remains secret).

To construct such a function, we may for example consider an \textit{authentication matrix} $ \mathcal{T} $, that is constructed with a \textit{strongly universal} keyed function $ h(\mathcal{K},\mathcal{M}) $ for the keyspace $ \mathcal{K} $ and an arbitrary input string $ \mathcal{M} $, i.e.  $ \mathcal{T} = \abs{\mathcal{K}}  \times  \abs{\mathcal{M}}  $. 
Typical examples for $ h $ were studied in~\cite{Wegman.1981,gilbert1974codes}, but it can also be a computationally secure MAC or HMAC function.

To generate an authentication tag for a message $ m \in \mathcal{M}$ and a given key $ k \in \mathcal{K} $, one takes the corresponding cell $ t \in \mathcal{T} $ as an out
put.
Assuming a uniform distribution of $ \mathcal{K} $ and generating a new $ \mathcal{T} $ for every message $ m \in \mathcal{M} $, the probability of forging a valid authentication tag is $p_t = 1/\abs{\mathcal{T}}$ if the key is only used once.

The tag space $\abs{\mathcal{T}}$ depends on the message- and key space.
For a message space of size $ \abs{\mathcal{K}} = \abs{\mathcal{M}}^2 $, the probability of forging a valid authentication tag is 
\[ p_t = \frac{1}{\abs{\mathcal{T}}} = \frac{\abs{\mathcal{M}}}{\abs{\mathcal{K}}} = \frac{1}{\sqrt{\abs{\mathcal{K}}}} \].

A construction as above is referred to \textit{1-time-secure}, whereas similar other \textit{n-time-secure} constructions exist if $ k \in \mathcal{K} $ is to be used multiple times~\cite{Fak.1979,rosenbaum1993lower}.

Note that, while assuming such a matrix is not necessary for ITS authentication, we use it as an example to simplify the above cheating probability derivation.

\section{Security against double-spending}\label{sec:quantumproof}

\subsection{For $N=1$.}\label{sec:nequal1}

This section derives the security analysis for a token consisting of $N=1$ quantum state. The aim is to derive a border between the secure region of operation, containing all pairs of experimental imperfections $\left(l,e\right)$ for which the presence of a malicious party can be detected, and its corresponding insecure region, containing all pairs of dishonest experimental deviations $\left(l,e\right)$ for which a malicious behavior cannot be detected. In both cases, $l$ denotes the fraction of quantum states from $\ket{P}$ that are declared as losses, while $e$ denotes the fraction of quantum states from $\ket{P}$ for which the declared measurement outcome disagrees with the classical description $(b,\mathcal{B})$.

In the simplest case, a successful attack consists in producing two cryptograms $\kappa_0$ and $\kappa_1$  for two distinct Merchants $M_0$ and $M_1$ that both pass the TTP's verification test. We note that in this two-merchant scenario, the information contained in the output of the HMAC function is one bit regardless of the actual length of the output. We may therefore reduce the commitments $M_0$ and $M_1$ to measurements of $\ket{P}$ in the $Z$ and $X$ bases, respectively.

In order to succeed in their optimal attack, the dishonest party may perform any general quantum operation on  $\ket{P}$, and replace all lossy and noisy channels by perfect ones. The TTP may then detect an attack only if their (potentially tampered with) measured noise and losses lie within the secure region of operation. We use SDP techniques from \ref{sec:sdp} to minimize the errors $e$ that the adversary must induce/declare while introducing at most $l$ losses. We optimize over the set of all possible CPTP maps $\{\Lambda\}$, that produce two classical cryptograms living in Hilbert space $\mathcal{H}_0\otimes\mathcal{H}_1$ from the original experimental quantum token state $\rho_P$ living in $\mathcal{H}_{P}$. The resulting secure/insecure regions of operation are shown in Fig. 4.a in the main text.

We note that $\mathcal{H}_0$ and $\mathcal{H}_1$ are $3$-dimensional Hilbert spaces spanned by classical answers $\{\ket{a_0},\ket{a_1},\ket{\varnothing}\}$, where $\ket{a_0}$ and $\ket{a_1}$ are orthonormal basis vectors indicating two possible classical answers (0 and 1 respectively), and  $\ket{\varnothing}$ is a third basis vector (orthogonal to the two others) indicating the declaration of a lost state. On the other hand, $\mathcal{H}_{P}$ is a $7$-dimensional Hilbert space spanned by $\{\ket{v},\ket{q_0},\ket{q_1},\ket{m_0},\ket{m_1},\ket{m_2},\ket{m_3}\}$, where $\ket{v}$ is the vacuum state, $\ket{q_0}$ and $\ket{q_1}$ span a qubit space, and $\ket{m_i}$ constitute the four orthogonal outcomes which materialize the four perfectly distinguishable states in the multiphoton subspace. Since the states produced by SPDC are of the form $\sum_{n=0}^{\infty}c_n\ket{n}_1\ket{n}_2$ in the $\{\ket{n}\}$ photon number basis~\cite{BR:NJP10}, this leaves the individual subsystems in states of the form $\sum_{n=0}^{\infty}\widetilde{c}_n\ket{n}\bra{n}$. Our four states may then be written as the following density matrices :
\begin{alignat*}{4}
    \sigma_0 &= p_0\ketbra{v}{v}&&+p_1\ketbra{+}{+}&&+p_m\ketbra{m_0}{m_0}\\
    \sigma_1 &= p_0\ketbra{v}{v}&&+p_1\ketbra{+i}{+i}&&+p_m\ketbra{m_1}{m_1}\\
    \sigma_2 &= p_0\ketbra{v}{v}&&+p_1\ketbra{-}{-}&&+p_m\ketbra{m_2}{m_2}\\
    \sigma_3 &= p_0\ketbra{v}{v}&&+p_1\ketbra{-i}{-i}&&+p_m\ketbra{m_3}{m_3},
\end{alignat*}
where $\ket{+}$,$\ket{+i}$, $\ket{-}$, $\ket{-i}$ are the usual $X$ and $Y$ eigenstates in the qubit space spanned by $\ket{q_i}$ and the photon number populations $p_n$ are estimated from our experiment. This allows to express the experimental quantum token state $\rho_P$ as:

\begin{equation}
    \rho_P =\frac{1}{4} \sum_{k=0}^3 \sigma_k.
\end{equation}
The probability $P_0$ that $\kappa_0$ does not pass the TTP's verification is then given by:

\begin{equation}
P_0 = \Tr\sum_{k=0}^3\left(\frac{1}{2}\ket{a_{k}^\perp}\bra{a_{k}^\perp}\otimes \mathbb{1}\right) \Lambda\left(\frac{1}{4}\sigma_k\right),
\end{equation} 
while the probability $P_1$ that $\kappa_1$ does not pass the TTP's verification reads:

\begin{equation}
P_1=\Tr\sum_{k=0}^3\left(\mathbb{1}\otimes\frac{1}{2}\ket{a_{k}^\perp}\bra{a_{k}^\perp}\right) \Lambda\left(\frac{1}{4}\sigma_k\right).
\end{equation}
where $\ket{a_{k}^\perp}$ is the wrong, orthogonal answer to $\ket{a_{k}}$. Using \eqqref{eq:choi}, we may rewrite these expressions as $P_0 =\Tr\left(E_0 J(\Lambda)\right)$ and $P_1 = \Tr\left(E_1 J(\Lambda)\right)$, where $E_0$ and $E_1$ are the \textit{error operators}:

\begin{equation}
\begin{aligned}
   E_0=&\frac{1}{4}\sum_{k=0}^{3}\frac{1}{2} \ket{a_{k}^\perp}\bra{a_{k}^\perp}\otimes\mathbb{1}\otimes\overline{\sigma_k}, \\
   E_1=&\frac{1}{4}\sum_{k=0}^{3}\mathbb{1}\otimes\frac{1}{2} \ket{a_{k}^\perp}\bra{a_{k}^\perp}\otimes \overline{\sigma_k} .
\end{aligned}
\end{equation}
where $\overline{\sigma_k}$ denotes the complex conjugate of $\sigma_k$. 
Following a similar reasoning, the probability that the dishonest party declares losses for $\kappa_0$ (resp. $1$) reads $\Tr\left(L_0 J(\Lambda)\right)$ (resp. $\Tr\left(L_1 J(\Lambda)\right)$), where $L_0$ and $L_1$ are the \textit{loss operators}, containing the projection onto the state $\ket{\varnothing}$:
\begin{equation}
\begin{aligned}
L_0 = \frac{1}{4}\sum_{k=0}^3 \ket{\varnothing}\bra{\varnothing}\otimes\mathbb{1} \otimes\overline{\sigma_k}, \\
L_1= \frac{1}{4}\sum_{k=0}^3
\mathbb{1}\otimes\ket{\varnothing}\bra{\varnothing} \otimes\overline{\sigma_k} .
\label{will}
\end{aligned}
\end{equation}
We now search for the optimal CPTP map $\Lambda$ that minimizes $e$ for a fixed $l$. We recast this problem as the following primal SDP (which we choose to be a minimization problem rather than a maximization problem for the sake of intuition):
\begin{equation}\label{eq:primal}
\begin{aligned}
\min\quad& \Tr\left(E_0 J(\Lambda) \right) \\
\text{s.t. }&  \Tr_{\mathcal{H}_0\otimes\mathcal{H}_1}\left(J(\Lambda)\right) = \mathbb{1}_{\mathcal{H}_P} \\
    &\Tr\left(E_0 J(\Lambda) \right) \geqslant \Tr\left(E_1 J(\Lambda) \right) \\
    &\Tr\left(L_0 J(\Lambda) \right) \leqslant l \\
    &\Tr\left(L_1 J(\Lambda) \right) \leqslant l  \\
    &J(\Lambda) \geqslant 0
\end{aligned}
\end{equation}
The first constraint imposes that $\Lambda$ is trace-preserving, the second imposes that the error rate for cryptogram $\kappa_0$ is at least equal to that for cryptogram $\kappa_1$, the third and fourth impose that the losses declared for $\kappa_0$ and $\kappa_1$ do not exceed the expected honest losses, and the fifth imposes that $\Lambda$ is completely positive. 

The numerical primal optimal values $\{e^{\left(\textnormal{primal},1\right)}\}$ of \eqqref{eq:primal} are plotted in Fig. 4.a in the main text. as a function of loss tolerance $l$. Following the methods from \ref{sec:sdp}, we derive the dual problem associated with \eqqref{eq:primal} to prove that $e^{\left(\textnormal{primal},1\right)}$ provides a tight upper bound on the cheating probability. This problem can be written as:

\begin{equation}
\begin{aligned}
    \textnormal{max } &[\mathrm{Tr}[X_1] + l \cdot (x_2+x_3)]\\
    \textnormal{s.t. } &[\mathbb{1}_9\otimes X_1 + L_0^\dagger \cdot x_2 + L_1^\dagger \cdot x_3 + (E_1-E_0)^\dagger \cdot x_4 - E_0^\dagger] \leq 0 \\
    &\textnormal{where }  X_1 \textnormal{ is a hermitian $2\times2$ matrix}\\
    & x_2, x_3, x_4 \in \mathbb{R},
\end{aligned}\label{eq:dual}
\end{equation}
and its numerical optimal dual value $e^{\left(\textnormal{dual},1\right)}$ indeed satisfies $e^{\left(\textnormal{primal},1\right)}=e^{\left(\textnormal{dual},1\right)}$. 
Note that the error and loss operators are hermitian, i.e. $L_x^\dagger = L_x$ and $E_x^\dagger = E_x$.

\subsection{For $N\rightarrow \infty$.}

In this section we show that, when $N\rightarrow\infty$, a malicious party does not gain any advantage in correlating the $N$ states in the quantum token (i.e., that the single-state security bounds derived in \ref{sec:nequal1} still hold). Following the exponential de Finetti arguments from \cite{Renner:Finetti}, it is sufficient to argue that, since our quantum token state is symmetric under arbitrary re-ordering of the $N$ quantum states, the individual states are well approximated by a mixture of independent and identically distributed states.

The security analysis based on semidefinite programs is convenient for $N=1$ quantum states, and for proving that the resulting cheating strategy is indeed optimal. In our particular case, one can also derive an analytical expression that fits the optimal cheating strategy derived in \ref{sec:nequal1}:
\begin{equation}
    e^{\left(\textnormal{primal},N\right)} =  e^{\left(\textnormal{dual},N\right)} = -\frac{1}{4+2\sqrt{2}}\cdot l + \frac{1-\frac{p_m}{2}}{8+4\sqrt{2}}
    \label{eq:Merr}
\end{equation}
which is a function of losses $l$ and multiphoton emission probability $p_m$. We can therefore easily determine how many errors a malicious Client has to introduce in order to comply to a certain constraint on losses and still double spend. Equivalently, through simple inversion of this equation, we can determine the amount of losses needed for successful cheating given the amount of declared errors $e$:
\begin{equation}
    l^{\left(\textnormal{primal},N\right)}=l^{\left(\textnormal{dual},N\right)}= \Big[-(4+2\sqrt{2})\cdot \frac{e}{1-\frac{p_m}{2}} +\frac{1}{2}\Big] \cdot (1-\frac{p_m}{2}) = -(4+2\sqrt{2})\cdot e +\frac{1}{2}\cdot (1-\frac{p_m}{2})
    \label{eq:Mloss}
\end{equation}

\eqqref{eq:Merr} and \eqqref{eq:Mloss} describe the secure region from Fig. 4.a in the main text. For simplicity, we define a new parameter $\mathcal{M}(e,l)$, which indicates the overall amount of \textit{mishaps} (i.e. any combination of errors and losses) a TTP might receive from a malicious party. By upper bounding this expression, we can easily specify the secure region:

\begin{equation}
    \mathcal{M}(e, l) = (4+2\sqrt{2})\cdot e + l \leq \frac{1}{2}\cdot(1-\frac{p_{m}}{2})
\end{equation}

\subsection{For $N$ finite}
Since $N$ will be finite in a realistic implementation, it is necessary to study the effect of finite-length statistics on the honest and dishonest success probabilities $p_h$ and $p_d$, respectively. A malicious party may indeed successfully cheat by introducing fewer losses or errors than the expected asymptotic values displayed in Fig. 4.a in the main text. We will make use of Chernoff-Hoeffding inequalities from \eqqref{Chernoff_Hoeffding} to bound this probability.

\subsubsection{Honest success probability.}
While the TTP allows for a certain amount of losses and errors, in order for the protocol to work in a realistic, i.e. imperfect scenario, there is still some probability, that the honest parties actually introduce more than the expected number of errors and/or losses, i.e. $\mathcal{M}(e_h^{\textnormal{act}},l_h^{\textnormal{act}})$. We denote the probability that this occurs as $p^{\textnormal{fail}}_h$. Following \eqqref{Chernoff_Hoeffding}, it is possible to upper bound $p^{\textnormal{fail}}_h$ as:
\begin{equation}\label{honestaborterrors}
    p^{\textnormal{fail}}_{h} = P\left[\mathcal{M}(e_h^{\textnormal{act}}, l_h^{\textnormal{act}})\geq (1+\delta_h)\cdot \mathcal{M}(e_h, l_h)\right] \leq \exp{\left(-\frac{(\delta_h)^2}{3}\cdot\mathcal{M}(e_h, l_h)\cdot N\right)}
\end{equation}
for some $\delta_h$. The honest success probability $p_h$ of the protocol is then defined as the probability that the protocol does \textit{not} abort when it is followed honestly. Therefore it can be expressed as:
\begin{equation}\label{correctness}
    p_h = 1- p_h^{\textnormal{fail}} \geq 1- \exp{\left(-\frac{(\delta_h)^2}{3}\cdot\mathcal{M}(e_h, l_h)\cdot N\right)}
\end{equation}

As is apparent, the correctness increases exponentially with $\delta_h^2$ and $N$. However, since the TTP also has to allow for more errors and losses with increasing $\delta_h$, they are more vulnerable to malicious parties. We thus need to ensure that:

\begin{equation}\label{eq:corrconstraint}
    \mathcal{M}(e_h, l_h) \cdot (1+\delta_h) < \frac{1}{2} \cdot\left(1-\frac{p_m}{2}\right).
\end{equation}
This inequality upper bounds the allowed value of $\delta_h$ without jeopardizing the information theoretical security of the protocol.
\medskip

\subsubsection{Dishonest success probability.}
Similarly, it might be possible for a cheating party to introduce fewer errors and losses than expected from the theoretical security proof. Following \eqqref{Chernoff_Hoeffding}, we upper bound this probability $p_d$ as:

\begin{equation}\label{cheatingChernoff_Hoeffding}
   p_d = P\left[\mathcal{M}(e^{\textnormal{act}}_d, l^{\textnormal{act}}_d) \leq (1-\delta_d)\cdot \frac{1}{2} \cdot(1-\frac{p_m}{2})\right] \leq \exp{\left(-\frac{(\delta_d)^2}{2}\cdot \frac{1}{2}\cdot(1-\frac{p_m}{2})\cdot N\right)}
\end{equation}
for some $\delta_d$. Using \eqqref{eq:corrconstraint}, we must then ensure that:

\begin{equation}
    \frac{1}{2}\cdot\left(\left(1-\frac{p_m}{2}\right)\cdot(1-\delta_d)\right) \geq \mathcal{M}(e_h, l_h)\cdot (1+\delta_h).
\end{equation}

Since this is the amount of mishaps the TTP allows for in order to assure the correctness of the protocol.

\section{Security of the Cryptogram}\label{sec:cryptogram}
In our protocol, we use an \textit{i.t.-secure MAC} to compute the measurement basis string for the quantum token $ \ket{P} $.
I.e., we measure $ \ket{P} $ according to $ \mac(C_i,M_i) $, where $ C_i \in C $ is a preshared key of the Client with the TTP, and $ M_i $ is the Merchant's ID where the payment token is spent.
This facilitates the security of the quantum channels as well as the classical channels.

To guarantee a decent level of security, one has to fare the number of quantum states $ N $ for a single payment token  (see \ref{sec:quantumproof}) with the number of merchants $ \abs{M} $ and the size of the output tag $ \abs{\mathcal{T}} $.
In the case of a \textit{1-time-secure} function, we assume 
\[ \forall t\in\mathcal{T} \text{ that } \abs{\mathcal{T}} = \sqrt{\abs{\mathcal{K}}} = \sqrt{\abs{C}} \ggg \abs{M} \] 
for a single authentication tag.
The probability $p_t$ of forging the output of $ \mac(C_i,M_i) $ should be similarly low as the \textit{dishonest success probability} $ p_d $ for a given sub-token size $ N $ (see \ref{sec:quantumproof}, Fig. 4 in the main text).\\

\noindent
Thus, we choose $ p_d \sim p_t \Rightarrow p_d \sim \frac{\abs{M}}{\sqrt{\abs{C}}} \sim \frac{1}{\sqrt{\abs{C}}}$, and the overall the token length  $\lambda=N \cdot \log_2{\abs{C}}$.

\noindent
We stress that this also holds for \textit{n-time-secure} constructions (see \ref{sec:hmac-it-secure}), just that $ \abs{C} $ needs to be chosen accordingly.

\section{Intuition about potential attacks on untrusted channels}

In the following we will discuss the potential attacks our protocol protects against, and en passant explain which part of our scheme serves which precise purpose. Note that the following is only an intuitive explanation of the rigorous security proof provided in the previous section.\\

\subsection{Compromising classical channels}
We have two classical channels in our protocol, namely \texttt{CH2} (Client $\rightarrow$ Merchant) and \texttt{CH3} (Merchant $\rightarrow$ TTP) in FIG.\,1 of the main text. Since both of them are untrusted, it is possible for a malicious third party to intercept them and modify the cryptogram $\kappa(C,M_i,\ket{P})$ towards another merchant $M_i'$ on \texttt{CH2} or change the merchant's Id $M_i$ towards another merchant's $M_i'$ on \texttt{CH3}.

\begin{description}
    \item[CH2] 
        To be accepted by the TTP, the attacker has to find another $\kappa'(C,M_i', \ket{P})$ for the Client's secret $C$ that commits the purchase to another Merchant $M_i'$. This is impossible for two reasons: \\[.5em]
        1.) the attacker would need to determine $C$ from the Client to calculate a second measurement bases $m' = \mac(C,M_i')$, which is supposed to be securely distributed between Client and TTP. It is impossible to determine $C$ from $\kappa$ as the function of the measurement basis $\mac(C,M_i)$ is information-securely irreversible -- and its output is additionally hidden in the quantum measurement and therefore unknown to the attacker. \\[.5em]
        2.) even if the attacker would have access to $C$ for some reason -- e.g. by accessing the Client's memory -- , he would require the classical description of quantum token $\ket{P}$, since it is already measured and quantum measurements are destructive. However, the classical description is only known by the TTP and never communicated.
    \item[CH3] 
        To change $M_i$ that is communicated together with $\kappa$, the attacker has to find $M_i'$ that generates the same measurement bases $m' = \mac(C,M_i')$ that was used to generate $\kappa$. To do so, he would need access $C$, which is supposed to be securely distributed between Client and TTP. However, even if he would have access, the chances of finding a collusion such that $\mac(C,M_i) = \mac(C,M_i')$ for a given $C$ are exponentially low due to the information-theoretic nature of the \mac{} function.\\[.5em]
        If the Merchant requires instant notification of the payments acceptance, however, this channel requires authentication s.t. the Client could not alter this message.
\end{description}
Please note, that both attacks can be performed by a malicious Merchant as well -- who has access to both channels and is supposed to be untrusted -- but fail for the same reason.

\subsection{Compromising the quantum channel}
A significant advantage of our scheme is that it is preferable but \textit{not necessary} to authenticate the quantum channel used to distribute $\ket{P}$. Let us suppose that a malicious party intercepts the quantum states $\ket{P}$ and sends their own quantum states $\ket{P'}$ to the Client instead.\\
After the Client measures $\ket{P'}$ in the basis $\mac(C,M_i)$, they will hold the cryptogram $\kappa' = \kappa(C,M_i,\ket{P'})$. If $\kappa'$ reaches the TTP, the transaction will be declined since $\kappa(C,M_i, \ket{P'}) \neq \kappa(C,M_i,\ket{P})$ (within the error/loss tolerance allowed by the security analysis). This means, that the Client as well as the TTP will be able to detect that the quantum states have been tampered with and that precautions should be taken.\\
Another possible cheating strategy is for the malicious party to use the quantum token $\ket{P}$ themself and measure it in another basis than the Client had intended. However, the malicious party does not know $C$, since it was securely distributed only between Client and TTP, and is thus unable to determine any measurement basis $m_j$ that will be accepted for the Merchant that they choose.\\

\subsection{Compromising both channels simultaneously}
Let us now suppose that a malicious party intercepts $\ket{P}$ on the quantum channel, replaces it with another quantum token $\ket{P'}$, and waits for the honest Client to send the resulting $\kappa'$ on the classical channel. If the Client would measure $\ket{P'}$ in a basis that is dependent on $C$ in a simple way, e.g. $m_i = M_i \oplus C$ then the malicious party gains knowledge of $C$, and can then substitute the Client's identity in multiple transactions. This is why we use a MAC instead: even if the malicious third party gets hold of $\kappa'$ and, by knowing $\ket{P'}$, deduces the measurement basis $m_i$, they are unable to retrieve $C$, because of the information theoretically secure nature of the used MAC, i.e. because the number of collision ensures that no cheating strategy would be better than guessing. Thus again, the TTP (and subsequently the Client) realise that something is wrong, while the secret Client token $C$ remains hidden. Depending on the nature of the MAC the token $C$ may resist a certain amount of failures, before it has to be exchanged.\\

\subsection{Experimental details}\label{sec:experiment}
This section is dedicated to our protocol performance and setup characterization. \tabref{tab:Losses} presents the measured transmission and corresponding loss rate of various setup components, while \tabref{tab:my_label} details the setup characterization and security performance. 

\begin{table}[h!]
    \centering
    \begin{tabular}{lcc}
      \hline\hline
         Component &  Transmission & Losses
           \\\hline
         Fibre link& 93.4 $\pm$ 1.5~\% & 6.9 $\pm$ 1.5~\%
           \\
         Client setup & 89.82 $\pm$ 0.57~\% & 10.18 $\pm$ 0.57~\%
           \\
         Detector 1 & 93.34 $\pm$ 0.75~\% & 6.66 $\pm$ 0.75~\%
           \\
         Detector 2 & 92.51$\pm$ 0.78~\% & 7.49 $\pm$ 0.78~\%
          \\ \hline
         Overall & 77.6 $\pm$ 1.5~\% & 22.4 $\pm$ 1.5~\%\\\hline\hline
    \end{tabular}
    \caption{\textbf{Transmission and corresponding losses of the experimental setup.}}
    \label{tab:Losses}
\end{table}

\begin{table}[h!]
    \centering
    \begin{tabular}{llcc}
    \hline\hline
    &Parameter name & Variable& Value \\\hline
    \parbox[t]{3mm}{\multirow{8}{*}{\rotatebox[origin=c]{90}{Experimental results}}}& Losses& $l_m$ & 22.4 $\pm$ 1.5~\% \\
         &Errors H/V& $e_m^{H/V}$ & 1.4491 $\pm$ 0.0083~\%\\
         &Errors +/- & $e_m^{+/-}$ & 3.278 $\pm$ 0.013~\%\\
         & Mishaps & $\mathcal{M}(e^{+/-}_m, l_m)$ & 0.448 $\pm$ 0.016 \\
         &Sub-token length & $N$ & 4 482 440 $\pm$ 1600\\
         & Q-bit rate & $1/s$ & 14 298 $\pm$ 6 Hz\\
         & Correlation function &$g_2(0)$ & 0.03010 $\pm$ 0.00014\\
         &Multiphoton emission &$p_m$ & $\leq$6.02 $\pm$ 0.28~\%\\
         \hline
         \parbox[t]{3mm}{\multirow{3}{*}{\rotatebox[origin=c]{90}{SDP}}}&Dishonest errors & $e_d(l_m)$ & 3.82 $\pm$ 0.22~\%\\
         &Dishonest losses H/V & $l_d(e_m^{H/V})$ &  38.600 $\pm$ 0.058~\%\\
         &Dishonest losses +/-& $l_d(e_m^{+/-})$ & 26.111 $\pm$ 0.090~\% \\
         \hline
         \parbox[t]{3mm}{\multirow{5}{*}{\rotatebox[origin=c]{90}{C-H bounds}}}& Allowed mishaps &$\mathcal{M}(e, l)$ & 0.4849 \\
         &Tolerance correctness & $\delta_h$ & 0.031 \\
         &Honest success probability & $p_h$ & $\sim 1$ \\
         & Tolerance cheating & $\delta_d$ & 0.01 \\
         &Dishonest success probability & $p_d$ &$5.911\pm0.088$ $\cdot 10^{-45}$\\
         \hline\hline
    \end{tabular}
    \caption{\textbf{Setup characterization and protocol performance.} 
    }
    \label{tab:my_label}
\end{table}

\bibliography{supplement}